# Optimal Capacity-Driven Design of Aperiodic Clustered Phased Arrays for Multi-User MIMO Communication Systems

N. Anselmi,[(1)] *Member, IEEE*, P. Rocca,[(1)(2)] *Senior Member, IEEE*, S. Feuchtinger,[(3)] B. Biscontini,[(3)] A. M. Barrera,[(3)] and A. Massa,[(1)(4)(5)] *Fellow, IEEE*

[(1)] *CNIT* - "University of Trento" Research Unit
Via Sommarive 9, 38123 Trento - Italy
E-mail: {*nicola.anselmi*, *paolo.*rocca, *andrea.massa*}@*unitn.it*
Website: *www.eledia.org/eledia-unitn*

[(2)] *ELEDIA Research Center* (*ELEDIA@XIDIAN* - Xidian University)
P.O. Box 191, No.2 South Tabai Road, 710071 Xi'an, Shaanxi Province - China
E-mail: *paolo.rocca@xidian.edu.cn*
Website: *www.eledia.org/eledia-xidian*

[(3)] *HUAWEI TECHNOLOGIES DUESSELDORF GmbH - MUNICH RESEARCH CENTER*
*Riesstrasse 25 C-2.0G, 80992 Munich - Germany*
E-mail: {*stefan.feuchtinger*, *bruno.biscontini*, *alejandro.murillo.barrera*}@*huawei.com*
Website: *www.huawei.com*

[(4)] *ELEDIA Research Center* (*ELEDIA@UESTC* - UESTC)
School of Electronic Engineering, Chengdu 611731 - China
E-mail: *andrea.massa@uestc.edu.cn*
Website: *www.eledia.org/eledia-uestc*

[(5)] *ELEDIA Research Center* (*ELEDIA@TSINGHUA* - Tsinghua University)
30 Shuangqing Rd, 100084 Haidian, Beijing - China
E-mail: *andrea.massa@tsinghua.edu.cn*
Website: *www.eledia.org/eledia-tsinghua*

# Optimal Capacity-Driven Design of Aperiodic Clustered Phased Arrays for Multi-User MIMO Communication Systems

N. Anselmi, P. Rocca, S. Feuchtinger, B. Biscontini, A. M. Barrera, and A. Massa

**Abstract**

The optimal design of aperiodic/irregular clustered phased arrays for base stations (*BS*s) in multi-user multiple-input multiple-output (*MU-MIMO*) communication systems is addressed. The paper proposes an *ad-hoc* synthesis method aimed at maximizing the users traffic capacity within the cell served by the *BS*, while guaranteeing the sufficient level of signal at the terminals. Towards this end, the search of the optimal aperiodic clustering is carried out through a customized tiling technique able to consider both single and multiple tile shapes as well as to assure the complete coverage of the antenna aperture for the maximization of the directivity. Representative results, from a wide set of numerical examples concerned with realistic antenna models and benchmark *3GPP* scenarios, are reported to assess the advantages of the irregular array architectures in comparison with regular/periodic layouts proposed by the standard development organizations, as well.

# 1 Introduction

Modern mobile communications rely on *MU-MIMO* systems to connect a growing number of wireless devices (e.g., smart-phones, internet-of-things (IoT) sensors, vehicles, etc ... [1][2]), and to assure a better quality-of-service (*QoS*) to the users (*UE*s), namely lower latency, higher throughput, and more resilient links [3][4]. These features are enabling factors for many interesting and innovative applications such as autonomous driving and remote surgery [5][6], just to mention a few. One fundamental enabler of such a wireless revolution is the phased array (*PA*) technology, which is used for the radio access to the network so that multiple *UE*s can be jointly served with high-quality and high-throughput links [7]-[9]. Indeed, the possibility to dynamically generate simultaneous beams pointing along different directions with desired beam shapes makes *PA*s a very attractive solution for the needs of nowadays and next generation wireless communications. However, conventional multi-beam *PA*s are too expensive for a large scale deployment of *BS*s [10] since they are characterized by beamforming networks with a huge number of radio-frequency (*RF*) chains that include transmission/receive modules (*TRM*s) with amplification/attenuation and time/phase delay stages. To reduce the costs, doing practicable the use of *PA*s technology, unconventional architectures have been widely studied and developed, originally, for radar and satellite applications [11] to minimize the number of *TRM*s through sparse (i.e., by arbitrary locating a minimum number of antennas over the array aperture) [12]-[14], thinned (e.g., by removing/switching-off a subset of the array elements, and the corresponding *TRM*s, of a fully-populated array) [15][16], and clustered (e.g., by grouping multiple elements into sub-arrays, each controlled by a single *TRM*) [17]-[22] arrangements of the array elements. Moreover, highly-modular clustered layouts, where the array is composed by simple modules/building blocks (i.e., tiles of contiguous elements) having the same structure [23]-[31] or belonging to a pre-defined set of tile shapes [32]-[38], have been more recently proposed [37] to further simplify the production and maintenance of *PA*s. In the framework of the *LTE* (long term evolution) and, more recently, of *5G* communications, clustered *PA* architectures have been proposed by the standard development organizations, such as the *3GPP* [39][40], to limit the number of *TRM*s. For instance, the baseline in [39] is a planar *PA* realized with vertical sub-arrays grouping an entire column of elements or part of them to better control the pattern shape along the azimuth plane. This is very useful to spatially separate the *UE*s in the cells when the relative position between the *BS* and the *UE*s is such that the horizontal beam scanning is uppermost required for guaranteeing high quality connections (e.g., the *BS* is located up on a tower and the *UE*s are on the ground or in lower buildings). On the contrary, the reduced number of *TRM*s in the columns of the *PA* makes the pattern control more challenging, if not impossible, due to the severe quantization of the amplitude and/or of the phase of the excitations in the vertical direction [41] when the beam has to be steered and/or shaped along the elevation plane.

Let us also consider that the number of *BS*s will tremendously grow with the advent of the *5G*, thus there will be many more scenarios where the beams generated from a *BS* will need both azimuth and elevation scanning and/or beam forming. Moreover, due to the increase of the operation frequencies, the objects in the environment will act more heavily as obstacles and impairments to the signal propagation, thus they must be carefully taken into account when designing the mobile wireless networks as well as the *BS*s. Furthermore, the limited costs and the low power consumption will be key performance indicators (*KPI*s) of the next generation of *PA*s for *BS*s.

In this framework, this paper proposes an innovative method for the synthesis of tiled *PA*s for next generation wireless communications. First, the modularity feature is targeted by considering a *PA* architecture composed by one or few simple tile shapes to reduce the manufacturing costs and to facilitate the maintenance towards a feasible mass-market production. Then, an aperiodic/irregular placement of the tiles within the *PA* aperture is done to yield the desired control of the beam pattern on the basis of the required horizontal and/or vertical scanning. To maximize the antenna efficiency and to profitably exploit all the available degrees of freedom (*DoF*s), a *PA* tiling algorithm that guarantees a perfect tessellation (i.e., a complete coverage) of the antenna aperture for *a-priori* given tile shape/s is used. Moreover, unlike standard approaches, which address the *PA* synthesis by neglecting the channel effects and optimize the free-space antenna *KPI*s (i.e., gain, sidelobe level, half-power beam-width, and no grating lobes condition [26][28][30][31][34][38]), a capacity-driven optimization strategy is here considered for the *BS* design in order to take into account the environment (i.e., the presence of obsta-cles/objects) for reaching the best communication performance [42]. To the best of the authors' knowledge, the main innovative contributions of this research work include (*i*) the design of clustered phased arrays for *BS*s through an *ad-hoc* strategy aimed at maximizing the traffic capacity of the users, while guaranteeing the coverage of the terminals within the cell, (*ii*) the development of a customized algorithm to define the optimal aperiodic tiling of the array elements, and (*iii*) the synthesis of innovative architectures for multi-user *MIMO* communication systems using either single or multiple tile shapes.

The rest of the paper is organized as follows. The *PA* synthesis problem at hand is mathematically described and formulated in Sect. 2, while the tiling algorithm is presented in Sect. 3. Section 4 deals with the numerical analysis and the comparative assessment of the proposed synthesis approach. Eventually, some conclusions and final remarks are drawn (Sect. 5).

## 2  Mathematical Formulation

Let us consider a mobile communication scenario partitioned into hexagonal cells, $d_H$ being the edge length, where each site is composed by three adjacent cells sharing an edge two by two (Fig. 1) and it has, at the center, a *BS* that mounts three *PA* panels at a height $h_{BS}$ above the ground. Moreover, the minimum distance between two *BS*s, the so-called inter-site distance (*ISD*), is equal to $ISD = 3 \times d_H$, while the *UE* terminals within a cell are served by a single *PA* consisting of a rectangular planar arrangement of $M \times N$ dual/cross-polarized radiating elements, $\psi$ ($\psi \in \{V, H\}$) being the polarization state of the *BS* antennas. The array elements are supposed to lie in the $(y, z)$-plane over a regular rectangular lattice and to be spaced by $d_y$ and $d_z$ along the $y$ and $z$ axis, respectively (Fig. 2). To reduce the complexity and the costs of the feeding network with respect to a fully-populated *PA* architecture[1], the $M \times N$ elements are grouped into $Q$ tiles ($Q < M \times N$), connected at sub-array level to two TRMs, one for each polarization. The membership of each $(m, n)$-th ($m = 1, ..., M$; $n = 1, ..., N$) element to a $q$-th ($q = 1, ..., Q$) sub-array is coded into the sub-array aggregation vector[2]

---

[1]In a fully-populated array, each antenna has a dedicated *TRM* for the $V$-polarization and another one for the $H$-polarization. Thus, the total number of *TRM*s turns out to be $2 \times (M \times N)$.
[2]Without loss of generality and hereinafter, the same clustering of the array elements is considered for the two polarizations (i.e., $\mathbf{s}_\psi = \mathbf{s}$, $\psi \in \{V, H\}$).

$$\mathbf{s} \triangleq \{s_{m,n} \in [1:Q]; m = 1, ..., M; n = 1, ..., N\}. \tag{1}$$

For the sake of example, Figure 3 shows two clustered configurations along with the integer-valued entries of the sub-array aggregation vector **s**. The former [Fig. 3(*a*)] is characterized by a baseline/regular tiling with vertical sub-arrays, while the other [Fig. 3(*b*)] is an irregular tiling of the same aperture where each color corresponds to a different tile orientation/flip.

By considering a linear combination of the sub-array outputs [43], a tiled *PA* generates up to $B = Q$ independent beams for each $\psi$-th ($\psi \in \{V, H\}$) polarization, the radiated $b$-th ($b = 1, ..., B$) $\psi$-th ($\psi \in \{V, H\}$) polarized electromagnetic (*EM*) field in the far-field zone being mathematically expressed as follows

$$\mathbf{E}^{(b,\psi)}(\theta, \phi) = \sum_{m=1}^{M} \sum_{n=1}^{N} \mathbf{e}_{m,n}^{(b,\psi)}(\theta, \phi) w_{m,n}^{(b,\psi)} e^{j\frac{2\pi}{\lambda}(y_m \sin\theta \sin\phi + z_n \cos\theta)} \tag{2}$$

where $\mathbf{e}_{m,n}^{(b,\psi)}(\theta, \phi)$ is the active element pattern of the $(m,n)$-th ($m = 1, ..., M; n = 1, ..., N$) $\psi$-polarized ($\psi \in \{V, H\}$) antenna centered at $\mathbf{r}_{m,n} = (0, y_m, z_n)$ with non-null coordinates equal to

$$y_m = (m - 1) \times d_y \tag{3}$$

and

$$z_n = h_{BS} + \left(n - \frac{N+1}{2}\right) \times d_z, \tag{4}$$

$\lambda$ and $(\theta, \phi)$ being the wavelength and the observation direction, respectively. In (2), the sets of $M \times N$ equivalent element excitations, $\mathbf{W} = \{\mathbf{w}^{(b,\psi)}; b = 1, ..., B; \psi \in \{V, H\}\}$ being $\mathbf{w}^{(b,\psi)} = \{w_{m,n}^{(b,\psi)}; m = 1, ..., M; n = 1, ..., N\}$, are function of the corresponding sets of $Q$ sub-array complex coefficients, $\mathbf{V} = \{\mathbf{v}^{(b,\psi)}; b = 1, ..., B; \psi \in \{V, H\}\}$ being $\mathbf{v}^{(b,\psi)} = \{v_q^{(b,\psi)}; q = 1, ..., Q\}$, according to the following relationship

$$w_{m,n}^{(b,\psi)} = \sum_{q=1}^{Q} \delta_{m,n,q} v_q^{(b,\psi)}, \tag{5}$$

where $\delta_{m,n,q}$ is the Kronecker delta function, which is equal to $\delta_{m,n,q} = 1$ if the $(m,n)$-th

($m = 1, ..., M$; $n = 1, ..., N$) array element belongs to the $q$-th ($q = 1, ..., Q$) sub-array (i.e., $s_{m,n} = q$) and $\delta_{m,n,q} = 0$, otherwise.

The $U$ *UE* terminals are randomly located within the hexagonal cell at $\{\mathbf{r}_u = (x_u, y_u, z_u);$ $u = 1, ..., U\}$ (Fig. 1) and they are equipped with dual-polarized receiving (RX) antennas, $\chi$ ($\chi \in \{V, H\}$) being the *RX* antenna polarization state. By assuming that the $b$-th ($b = 1, ..., B$) $\psi$-polarized ($\psi \in \{V, H\}$) beam generated by the *PA* is used to transmit a signal[3] to be received by the $\chi$-polarized ($\chi \in \{V, H\}$) antenna of the $u$-th ($u = 1, ..., U$) *UE* terminal (i.e., $U = Q$), the total power at the receiver is given by

$$\Pi^{(\chi)}(\mathbf{r}_u) = \Pi^{(\chi)}_{des}(\mathbf{r}_u) + \Pi^{(\chi)}_{mui}(\mathbf{r}_u) + \Pi^{(\chi)}_n(\mathbf{r}_u) \tag{6}$$

where $\Pi^{(\chi)}_{des}(\mathbf{r}_u)$ is the power associated to the desired signal, $\Pi^{(\chi)}_{mui}(\mathbf{r}_u)$ is the interference caused by the presence of the other ($U - 1$) *UE*s, and $\Pi^{(\chi)}_n(\mathbf{r}_u)$ is the power of the noise. More in detail, the desired power is given by

$$\Pi^{(\chi)}_{des}(\mathbf{r}_u) = \frac{\Psi}{B} \left| \sum_{m=1}^{M} \sum_{n=1}^{N} w_{m,n}^{(b,\psi)} \mathbf{G}^{(\psi,\chi)}(\omega; \mathbf{r}_u, \mathbf{r}_{m,n}) \right|^2 \tag{7}$$

where $\frac{\Psi}{B}$ is the average power radiated by the *PA*, which is supposed to be equal for all $B$ beams, $\Psi$ being the total power at the *PA* input, while $\mathbf{G}^{(\psi,\chi)}(\omega; \mathbf{r}_u, \mathbf{r}_{m,n})$ is the Fourier transform of the time-domain Green's function (*GF*) modelling the propagation and any other *EM* effect (e.g., the interactions of the transmitted signal with the objects/scatterers present in the environment, the mutual-coupling among the radiating elements, etc ...) between the $\psi$-polarized ($\psi \in \{V, H\}$) field generated by the ($m, n$)-th ($m = 1, ..., M$; $n = 1, ..., N$) element of the *PA* and the $\chi$-polarized ($\chi \in \{V, H\}$) *RX* antenna of the $u$-th ($u = 1, ..., U$) *UE* terminal [44][45][46]. Moreover, the multi-user interference power is equal to

$$\Pi^{(\chi)}_{mui}(\mathbf{r}_u) = \frac{\Psi}{B} \left| \sum_{o=1, o \neq b}^{B} \sum_{m=1}^{M} \sum_{n=1}^{N} w_{m,n}^{(o,\psi)} \mathbf{G}^{(\psi,\chi)}(\omega; \mathbf{r}_u, \mathbf{r}_{m,n}) \right|^2 \tag{8}$$

---

[3]Without loss of generality for the applicability of proposed design methodology, the transmitted signal is assumed to be monochromatic with carrier angular frequency $\omega$ ($\omega \triangleq 2\pi f$, $f$ being the working frequency) and characterized by a single pulse with unitary amplitude.

to account for the power transmitted at the $u$-th ($u = 1, ..., U$) *UE* in the position $\mathbf{r}_u$ and over the $\chi$ ($\chi \in \{V, H\}$) receiving polarization of its terminal, from the other ($B-1$) beams generated by the *PA* when feeding the sub-array *TRM*s with the coefficients $\mathbf{V}' \triangleq \{\mathbf{v}^{(o,\psi)}; o = 1, ..., B; o \neq b; \psi \in \{V, H\}\}$. As for the noise power, it is assumed equal to $\Pi_n^{(\chi)}(\mathbf{r}_u) = \sigma^2$ for all *UE* terminals as well as polarization states. Accordingly, the signal to interference-plus-noise ratio (*SINR*) associated to the signal received by the $\chi$-polarized ($\chi \in \{V, H\}$) antenna of the $u$-th ($u = 1, ..., U$) *UE* terminal, under the hypothesis of mutual incoherence between the signals transmitted through the $B$ beams and the noise [47], which is defined as $SINR^{(\chi)}(\mathbf{r}_u) \triangleq \frac{\Pi_{des}^{(\chi)}(\mathbf{r}_u)}{\Pi_{mui}^{(\chi)}(\mathbf{r}_u) + \Pi_n^{(\chi)}(\mathbf{r}_u)}$, turns out to be

$$SINR^{(\chi)}(\mathbf{r}_u) = \frac{\left|\sum_{m=1}^{M}\sum_{n=1}^{N} w_{m,n}^{(b,\psi)} \mathbf{G}^{(\psi,\chi)}(\omega; \mathbf{r}_u, \mathbf{r}_{m,n})\right|^2}{\left|\sum_{o=1, o\neq b}^{B}\sum_{m=1}^{M}\sum_{n=1}^{N} w_{m,n}^{(o,\psi)} \mathbf{G}^{(\psi,\chi)}(\omega; \mathbf{r}_u, \mathbf{r}_{m,n})\right|^2 + \frac{B\sigma^2}{\Psi}}. \quad (9)$$

Finally, the total *MU-MIMO sum-rate capacity* of the *PA* in downlink is given by [48]

$$C = \sum_{u=1}^{U} \sum_{\chi=V,H} C^{(\chi)}(\mathbf{r}_u), \quad (10)$$

where $C^{(\chi)}(\mathbf{r}_u)$ is the throughput capacity of the $\chi$-polarization ($\chi \in \{V, H\}$) of the $u$-th ($u = 1, ..., U$) *UE* [42]

$$C^{(\chi)}(\mathbf{r}_u) = \log_2\{1 + SINR^{(\chi)}(\mathbf{r}_u)\}. \quad (11)$$

To give statistically reliable evaluation of the down-link sum-rate capacity $C$ subject to a coverage of the hexagonal cell, the positions of the $U$ *UE*s, $\{\mathbf{r}_{u_p}; u_p = 1, ..., U\}$, are varied over $P$ different random configurations ($p = 1, ..., P$). The problem of synthesizing an irregular tiled array for *MU-MIMO* communication systems is then formulated as follows:

> *Capacity-Driven Tiled Array (CDTA) Synthesis Problem* - Given a *PA* of $M \times N$ dual-polarized radiating elements centered at $\{\mathbf{r}_{m,n} = (0, y_m, z_n); m = 1, ..., M; n = 1, ..., N\}$ and $P$ different configurations ($p = 1, ..., P$) of $U$ *UE*s randomly located at the positions $\{\mathbf{r}_{u_p}; u_p = 1, ..., U\}$ within the hexagonal cell, determine the optimal clustering of the array elements, $\widetilde{\mathbf{s}}$, by using $Q$ sub-arrays chosen among a finite set of $F$ tile shapes and the sets of sub-array complex coefficients for each $p$-th ($p = 1, ..., P$) *UEs* configuration, $\widetilde{\mathbf{v}}_p^{(b,\psi)} = \left\{ \widetilde{v}_{p,q}^{(b,\psi)}; q = 1, ..., Q \right\}$ ($b = 1, ..., B$; $\psi \in \{V, H\}$) such that the average sum-rate capacity

$$\widehat{C} = \frac{\sum_{p=1}^{P} \left\{ \sum_{u_p=1}^{U} \sum_{\chi=V,H} C^{(\chi)} \left( \mathbf{r}_{u_p} \right) \right\}}{P} \qquad (12)$$

is maximum and the power of the desired signal at the receiver position, $\mathbf{r}_{u_p}$ ($u_p = 1, ..., U$; $p = 1, ..., P$), for the $\chi$-polarization ($\chi \in \{V, H\}$) is above a minimum threshold $\Pi_{th}$, namely

$$\Pi_{des}^{(\chi)} \left( \mathbf{r}_{u_p} \right) \geq \Pi_{th}, \qquad (13)$$

to guarantee the connection of all UEs within the communication cell.

In order to address such a design problem, an innovative strategy, based on a customized optimization approach, is described in the following section (Sect. 3).

## 3 *CDTA* Design Method

The method for solving the *CDTA Synthesis Problem* consists of two nested loops and a solution selection step.

The outer loop exploits the Algorithm-X [49] to generate the sequence of all $T$ tiling configurations, $\{\mathbf{s}_t; t = 1, ..., T\}$, fully covering the *PA* aperture (*Tiling Problem*). Towards this end, the following notation is stated. Let us suppose that each $(m, n)$-th ($m = 1, ..., M$; $n = 1, ..., N$) array element is included into a square pixel such that the union of the $M \times N$ pixels defines an area $\mathcal{A}$ corresponding to the PA antenna aperture. Moreover, let us consider a set of tiles arrangements $\mathcal{T} \triangleq \{\tau_k; k = 1, ..., K\}$, where each $k$-th entry, $\tau_k$, corresponds to a $f$-th ($f = 1, ..., F$) polyomino (i.e., a planar figure composed of two or more pixels sharing at least one edge) located in $\mathcal{A}$ with an admissible (i.e., the polyomino is fully included in $\mathcal{A}$) position and orientation/flip. As an example, Figure 4 shows all the $K = 7$ positions and orientations (i.e., horizontal and vertical) that a domino tile (i.e., $F = 1$) can assume within $\mathcal{A}$. A trial $t$-th ($t = 1, ..., T$) tiling $\mathbf{s}_t$ is generated when a subset of $Q$ tiles $\widehat{\mathcal{T}}_t \triangleq \{\tau_{q_t} \in \mathcal{T}; q_t = 1, ..., Q\}$ covers all pixels of $\mathcal{A}$ without any overlap (i.e, $\bigcup_{q_t=1}^{Q} \tau_{q_t} = \mathcal{A}$ and $\bigcap_{q_t=1}^{Q} \tau_{q_t} = \emptyset$, $\emptyset$ being the empty set). With reference to the set of tiles arrangements $\mathcal{T}$ in Fig. 4, an admissible tiling of the aperture $\mathcal{A}$ is shown in Fig. 5($a$) (i.e., $\mathbf{s}_t = \{1; 1; 2; 3; 3; 2\}$,[(4)] $\widehat{\mathcal{T}}_t = \{\tau_1, \tau_3, \tau_7\}$). Given the aperture $\mathcal{A}$ and the whole set $\mathcal{T}$ of arrangements of the $F$ polyomino tiles, the binary incidence matrix $\mathbf{L} \triangleq \{l_{k,i} = \{0, 1\}; k = 1, ..., K; i = 1, ..., I\}$ ($I \triangleq M \times N$) is defined to encode whether the $i$-th ($i = 1, ..., I$; $i = m + (n-1) \times M$; $m = 1, ..., M$; $n = 1, ..., N$) pixel of $\mathcal{A}$ is either covered or not by the $k$-th ($k = 1, ..., K$) tiles arrangement $\tau_k$. The ($k,i$)-th entry of $\mathbf{L}$ is equal to one (i.e., $l_{k,i} = 1$) if the $i$-th ($i = 1, ..., I$) pixel of $\mathcal{A}$ is included into the $k$-th ($k = 1, ..., K$) entry of $\mathcal{T}$, otherwise it is equal to zero (i.e., $l_{k,i} = 0$). Figure 5($b$) shows the matrix $\mathbf{L}$ for the tiles arrangements in Fig. 4 where the rows related to the $Q = 3$ dominoes, which exactly tile $\mathcal{A}$ [Fig. 5($a$)], are highlighted with the corresponding colors. As it can be observed (Fig. 5), a tiling [e.g., Fig. 5($a$)] is obtained by selecting $Q$ rows of $\mathbf{L}$ such that a single "1" bit is present in all columns ($i = 1, ..., I$) for the $Q$ selected rows (i.e., all the pixels are covered without overlap). With reference to Fig. 5($b$), the tile $\tau_k|_{k=1}$ covers the pixels $i = 1$ and $i = 2$ (i.e., $l_{1,1} = 1$ and $l_{1,2} = 1$), the tile $\tau_k|_{k=3}$ covers the pixels $i = 4$ and $i = 5$ (i.e., $l_{3,4} = 1$ and $l_{3,5} = 1$), and the tile $\tau_k|_{k=7}$ covers the pixels $i = 3$ and $i = 6$ (i.e., $l_{7,3} = 1$ and $l_{7,6} = 1$). On the contrary, if the $Q$ rows have multiple "1" bit occurrences for one column/pixel [e.g., Fig. 5($d$)], two or more tiles overlap in that pixel, thus the corresponding tiled configuration [Fig. 5($c$)] turns out to be unfeasible/forbidden.

The inner loop is based on the Zero-Forcing (*ZF*) [10][48][50] to analytically compute [43][48] the sets of sub-array coefficients, $\mathbf{V}_{p,t}$ ($\mathbf{V}_{p,t} \triangleq \{\mathbf{v}_{p,t}^{(b,\psi)}; b = 1, ..., B; \psi \in \{V, H\}\}$), for each $t$-th ($t = 1, ..., T$) tiling and $p$-th ($p = 1, ..., P$) *UE*s placement so that the down-link average sum-rate capacity (12) is maximized (*Weighting Problem*). Indeed, the *ZF* coefficients maximize the *SINR* (9), which is directly proportional to the capacity (11), by minimizing the multi-user interference $\Pi_{mui}^{(\chi)}$ at the denominator of (9) placing nulls along the direction of all *UE*s except the desired one.

The optimal solution, $\widetilde{\mathbf{s}}$ and $\widetilde{\mathbf{V}}_p$ ($p = 1, ..., P$), is then selected among the $T$ ones generated by the two-loop process as the one that maximizes (12)

---

[(4)]For the sake of notation, $\mathbf{s}_t = \{s_{t,i}; i = 1, ..., I\}$ being $I = M \times N$ and $i = m + (n-1) \times M$ ($m = 1, ..., M$; $n = 1, ..., N$).

$$\left(\widetilde{\mathbf{s}},\,\widetilde{\mathbf{V}}_p\right) = \arg \max_{t=1,...,T} \left\{\widehat{C}\left(\mathbf{s}_t,\,\mathbf{V}_{p,t}\right)\right\} \qquad (14)$$

by jointly fulfilling the coverage condition (13)

$$\min_{u_p=1,...,U;\,p=1,...,P} \left\{\Pi_{des}^{(\chi)}\left(\mathbf{r}_{u_p};\,\mathbf{s}_t,\,\mathbf{V}_{p,t}\right)\right\} \geq \Pi_{th}. \qquad (15)$$

Such a three-step procedure for the solution of *CDTA Synthesis Problem* is implemented according to the following iterative step-by-step procedure:

- **Step 0 (*Tiling Problem*)** - *Incidence Matrix Definition.* Given the *PA* aperture $\mathcal{A}$ of $I = M \times N$ elements/pixels and $F$ polyomino tile shapes, identify all admissible positions and orientations/flips of each $f$-th ($f = 1, ..., F$) tile within $\mathcal{A}$ to fill the whole alphabet of tiles arrangements, $\mathcal{T} \triangleq \{\tau_k;\,k = 1, ..., K\}$. Define the incidence matrix $\mathbf{L}$ by adding a row of binary entries for each $k$-th occurrence of $\mathcal{T}$, $\tau_k$ ($k = 1, ..., K$), setting either $l_{k,i} = 1$ or $l_{k,i} = 0$ if the $i$-th ($i = 1, ..., I$) pixel is included or not into the $k$-th tile arrangement, respectively. Set $t = 1$, $q_t = 0$ and generate the sequence of tiling configurations $\mathbf{s}_t$ ($t = 1, ..., T$) fully/exactly covering $\mathcal{A}$ by means of the following recursive strategy:

- **Step 1 (*Tiling Problem*)** - *Pixel Selection.* Select the column of $\mathbf{L}$, $\mathbf{J}_{i^*} \triangleq \{l_{k,i^*};\,k = 1, ..., K\}$, $i^* \in [1,\,I]$ having the minimum number of "1" bits. If multiple columns have the same number of "1" bits, select the column having the lowest $i$ index [e.g., Fig. 6(*a*) - $i^* = 1$];

- **Step 2 (*Tiling Problem*)** - *Tile Placement.* Among the rows of $\mathbf{J}_{i^*}$ containing a "1" bit, select the one, $\mathbf{I}_{k^*} \triangleq \{l_{k^*,i};\,i = 1, ..., I\}$, $k^* \in [1,\,K]$ with the lowest $k$ index [e.g., Fig. 6(*a*) - $k^* = 1$]. In case this row has been previously selected, select the one with the second lowest $k$ index and proceed in this way until a new tile row is selected [e.g., Fig. 7(*c*) - $k^* = 5$]. In case all rows have been already selected, go to Step 7. Otherwise, mark as *placed* the $k^*$-th row of $\mathbf{L}$ [Fig. 6(*a*) and Fig. 7(*c*)], update $q_t$ ($q_t \leftarrow q_t + 1$), and place the tile $\tau_{k^*}$ in the aperture $\mathcal{A}$, $\tau_{q_t}|_{q_t=1} = \tau_{k^*}$ [Fig. 6(*b*) and Fig. 7(*d*)];

- **Step 3 (*Tiling Problem*)** - *Overlap Avoidance.* Mark as *covered* [e.g., Fig. 6(*a*) - gray

color] all the columns that in the $k^*$-th row, selected at Step 2, have a "1" bit [e.g., Fig. 6($a$) - columns $\mathbf{J}_1$ and $\mathbf{J}_2$] such that the corresponding pixels in $\mathcal{A}$ cannot be selected/covered again from another tile. Mark as *inadmissible* [e.g., Fig. 6($a$) - gray color] all rows of $\mathbf{L}$ having a "1" bit in the columns marked as covered [e.g., Fig. 6($a$) - $k = \{2, 5, 6\}$]. Remove from the current incident matrix, $\mathbf{L}$, all rows and columns marked as inadmissible and covered to yield $\widehat{\mathbf{L}}$;

- **Step 4 (*Tiling Problem*) -** *Admissible Tiling Check*. If $q_t < Q$ and $\widehat{\mathbf{L}}$ is not empty, update the incident matrix, $\mathbf{L} \leftarrow \widehat{\mathbf{L}}$, and go to Step 1 to place a new tile in $\mathcal{A}$ [e.g., Figs. 6($c$)-($d$)]. If $q_t = Q$ (i.e., all columns, $\{\mathbf{J}_i;\ i = 1, ..., I\}$, have been marked as covered) [e.g., Figs. 6($e$)-($f$)], set the sub-array aggregation vector $\mathbf{s}_t$ (Fig. 3) and go to Step 5 for computing the sub-array weighting coefficients. If $\widehat{\mathbf{L}}$ is empty, but $q_t < Q$ (i.e., all rows, $\{\mathbf{I}_k;\ k = 1, ..., K\}$, have been marked as inadmissible, but there are pixels/columns not covered), discharge the tiling at hand since unfeasible. Go to Step 7;

  - **Step 5 (*Weighting Problem*) -** *CDTA Weights Computation*. Compute the $p$-th ($p = 1, ..., P$) set of sub-array coefficients $\mathbf{V}_{p,t}$ by performing the Moore-Penrose pseudo-inversion of the concatenated *GF* channel matrix $\widehat{\mathbf{G}}_{p,t}$ ($\widehat{\mathbf{G}}_{p,t} \triangleq \{\mathbf{G}^{(\psi,\chi)}(\omega;\ \mathbf{r}_{u_p},\ \mathbf{r}_{m,n});$ $u_p = 1, ..., U;\ m = 1, ..., M;\ n = 1, ..., N;\ \psi \in \{V, H\}\}$) [42]

$$\mathbf{V}_{p,t} = \left[\widehat{\mathbf{G}}_{p,t}\right]^* \left(\widehat{\mathbf{G}}_{p,t} \left[\widehat{\mathbf{G}}_{p,t}\right]^*\right)^{-1} \tag{16}$$

  where $\cdot^*$ stands for conjugate transpose;

  - **Step 6 (*Weighting Problem*) -** *Cost Function Evaluation*. Given $\mathbf{s}_t$ and $\mathbf{V}_{p,t}$ ($p = 1, ..., P$) compute the average sum-rate capacity (12), $\widehat{C}_t = \widehat{C}(\mathbf{s}_t, \mathbf{V}_{p,t})$ and jointly check whether the coverage condition (13) holds true. If $\widehat{C}_t > \widetilde{C}$ then update the optimal solution $\left(\widetilde{\mathbf{s}}, \widetilde{\mathbf{V}}_p\right) \leftarrow (\mathbf{s}_t, \mathbf{V}_{p,t})$ and go to Step 7;

- **Step 7 (*Tiling Problem*) -** *Successive Tiling*. Remove the last placed tile $\tau_{k^*}$ from the aperture $\mathcal{A}$ and re-introduce in $\mathbf{L}$ the corresponding row [e.g., Fig. 7($a$) - $k^* = 7$ and Fig. 7($b$) - $k^* = 3$] as well as the corresponding rows and columns previously marked as inadmissible and covered. If $q_t = Q$ then increase the tiling index ($t \leftarrow t + 1$). Update $q_t$ ($q_t \leftarrow q_t - 1$) and go to Step 1.

## 4 Numerical Assessment

The objective of this section is twofold. On the one hand, the assessment of the proposed *CDTA Design Method* (Sect. 3) in synthesizing optimal tiling solutions affording maximum capacity, while guaranteeing the cell coverage. On the other hand, a critical evaluation on the effectiveness of aperiodic/irregular tiled *PA*s for realistic *MU-MIMO* communications defined according to the *3GPP* standards. As reference benchmark, a *PA* having $M = 8$ columns of $N = 12$ elements, $d_y = 0.5\lambda$ and $d_z = 0.7\lambda$ being the inter-element spacings, has been considered. Moreover, the dual-polarized slant $\pm 45$ [deg] ($H = +45$, $V = -45$) multi-layer stacked patch antenna in Fig. 8(*a*) has been chosen as radiating element and the corresponding element pattern within the array, $\mathbf{e}_{m,n}^{(b,\psi)}(\theta, \phi)$ ($m = 1, ..., M$, $n = 1, ..., N$, $\psi \in \{V, H\}$), has been determined with a full-wave simulation of the whole *PA* with the *ANSYS HFSS* software. Furthermore, the working frequency and the total *TX* power of the *BS* have been set to $f = 3.5$ [GHz] and $\Psi = 43$ [dBm], respectively. The *3D* propagation environment, which is mathematically described by the corresponding *GF*, has been numerically modeled with a geometry-based stochastic method that exploits an approximated ray-tracing approach based on the *QuaDRiGa* implementation [51][52][53], while the coverage threshold for the *UE* terminals has been fixed at $\Pi_{th} = -120$ [dBm]. As for the comparative assessment between regularly- and irregularly-clustered *PA*s, it has been carried out by considering as baseline a sub-arrayed array composed by $Q = 16$ vertical clusters, each containing $\frac{M \times N}{Q} = 6$ elements (i.e., an hexomino tile) [Fig. 8(*b*)], while two types of irregular hexominoes, namely the *P*-shaped and the *L*-shaped hexominoes (Fig. 9), have been selected for the aperiodic tiling.

The first scenario (*2D-UMa-O*) refers to a cell defined according to the *3GPP Urban Macro* (*UMa*) line-of-sight (*LOS*) model with $ISD = 500$ [m] (i.e., $d_H = 166, 67$ [m]) and illuminated by an outdoor *BS* placed at $h_{BS} = 25$ [m] above the ground (Tab. I). The number of *UE*s has been chosen equal to $U = Q = 16$ so that the *ZF* method could place nulls along the directions of all *UE*s except the desired one. The *UE*s have been randomly-distributed within the hexagonal cell at the ground floor (i.e., the *UE* height has been set to $h_{UE} = 1.5$ [m] - Tab. I) outside the buildings. More specifically, $P = 200$ configurations of the *UE*s positions, $\{\mathbf{r}_{u_p}; u_p = 1, ..., U; p = 1, ..., P\}$, have been defined as follows

$$\mathbf{r}_{u_p} = \{(x_{u_p}, y_{u_p}); x_{u_p} = c_{H,x} + R_{u_p} \cos(\alpha_{u_p}); \\ y_{u_p} = c_{H,y} + R_{u_p} \sin(\alpha_{u_p})\}, \quad (17)$$

where $R_{u_p}$ and $\alpha_{u_p}$ are real values uniformly randomly-selected within the intervals $0 \leq R_{u_p} \leq d_H$ and $0 \leq \alpha_{u_p} \leq 360$ [deg], respectively, while $\mathbf{c}_H = (c_{H,x}, c_{H,y})$ is the centroid of the hexagonal cell (Fig. 1). If a generic *UE* position, $\mathbf{r}_{u_p}$ ($u_p = 1, ..., U; p = 1, ..., P$), falls outside the hexagonal cell, its coordinates have been randomly re-generated until $U$ admissible *UE*s locations have been determined. For illustrative purposes, the whole set of $P \times U = 3200$ *UE*s of the *2D-UMa-O* scenario at hand are shown in Fig. 10(*a*) to make clear that the *UE*s positions cover the entire cell to allow a meaningful statistical assessment of the *PA* performance.

By considering *P*-shape hexomino tiles [Fig. 9(*a*) - $F = 1$], $T = 85926$ complete (i.e., fully and exactly covering the *PA* aperture) tiling configurations have been generated with the enumerative approach based on the Algorithm-X. The definition of the whole set of tiling configurations $\mathbf{s}_t$ ($t = 1, ..., T$) and their performance, in terms of both average sum-rate capacity (12) and coverage (13), have required 16 days and 11 hours of simulation on a standard single (i.e., no parallel computing has been exploited) *CPU* laptop.[5] The values of the average sum-rate capacity, ordered from the lowest to the highest, of the $T$ irregular *P*-shaped aperture tessellations, $\{\widehat{C}_t; t = 1, ..., T\}$, are reported in Fig. 11(*a*) and compared to the performance of the baseline/regularly-clustered configuration. As it can be observed, several solutions (i.e., $11.83\% \rightarrow 10165$ different tilings) perform better than the baseline, thus offering the antenna engineers a huge set of possibilities to select the most suitable *PA* architecture also taking into account additional design constraints and specifications. The optimal layout (i.e., the one maximizing the capacity $\widehat{C}$) is displayed in Fig. 10(*b*) by using a different color for each tile orientation/flip according to the notation of Fig. 9(*a*). Such a solution, $\widetilde{\mathbf{s}}$, outperforms the baseline in terms of average sum-rate capacity ($\widetilde{C} = 130.38$ [bps/Hz] vs. $\widehat{C}_{bsl} = 116.42$ [bps/Hz] - Tab. II) with an improvement of $\Delta\widehat{C} = 11.99\%$ ($\Delta\widehat{C} \triangleq \frac{\widetilde{C} - \widehat{C}_{bsl}}{\widehat{C}_{bsl}}$) (Tab. III). As for the coverage, Figure 11(c) plots the minimum values among the $P$ configurations of the desired power at the $a$-th ($a = 1, ..., A$; , $a = 2 \times (u_p - 1) + \mathcal{O}(\chi)$; $A \triangleq 2 \times U$) receiving port of the $U$ UEs, $\eta_{des}^{(a)}$ [$\eta_{des}^{(a)} \triangleq \min_{p=1,...,P} \left\{ \Pi_{des}^{(\chi)}(\mathbf{r}_{u_p}) \right\}$] in correspondence with the optimal tiling $\widetilde{s}$ in Fig. 10(b). One can notice that the minimum power level of both the baseline [Fig. 11(c)] and the optimized irregular layout are above the threshold $\Pi_{th}$ for the link connection, even though $\eta_{des}^{(a)}$ from the $P$-tiling is on average greater as confirmed by the statistics in Tab. II (e.g., $\eta_{des}^{min} \rfloor_{P-Shape} = -75.93$ [dBm] vs. $\eta_{des}^{min} \rfloor_{bsl} = -88.98$ [dBm] and $\eta_{des}^{avg} \rfloor_{P-Shape} = -68.27$ [dBm] vs. $\eta_{des}^{avg} \rfloor_{bsl} = -75.28$ [dBm], being $\eta_{des}^{min} \triangleq \min_{a=1,...,A} \left\{ \eta_{des}^{(a)} \right\}$ and $\eta_{des}^{avg} \triangleq \frac{1}{A} \times \sum_{a=1}^{A} \eta_{des}^{(a)}$).

---

[5]The standard laptop was equipped with an Intel-i5 CPU @ 1.60 GHz and 8Gb of RAM.

To further highlight the improved performance of the synthesized irregular layout in Fig. 10(b) as compared to the regular/baseline one [Fig. 8(b)], the cumulative distribution function (*CDF*) of the capacity performance,

$$CDF(C) = \int_{C_{\min}}^{C} PDF(\gamma) \, d\gamma, \qquad (18)$$

has been computed, the probability density function (*PDF*), which is proportional to the number of *UE*s served with a capacity value $\gamma$, being defined as

$$PDF(C) = Pr\left\{\gamma_1 \leq C_{u_p} < \gamma_2; u_p = 1, ..., U; p = 1, ..., P\right\} \qquad (19)$$

where $C_{u_p} = \sum_{\chi=V,H} C^{(\chi)}(\mathbf{r}_{u_p})$ is the total sum-rate capacity of the $u$-th ($u = 1, ..., U$) *UE* in the $p$-th ($p = 1, ..., P$) configuration. Moreover, $\gamma_1 = C_{\min} + (\upsilon - 1) \times \delta C$ and $\gamma_2 = C_{\min} + \upsilon \times \delta C$ ($\upsilon = 1, ..., \Upsilon$), $\delta C = \frac{C_{\max} - C_{\min}}{\Upsilon}$, and with $\Upsilon = 20$, while $C_{\min}$ and $C_{\max}$ are the minimum ($C_{\min} \triangleq \min_{u_p=1,...,U; p=1,...,P} \{C_{u_p}\}$) and the maximum ($C_{\max} = \max_{u_p=1,...,U; p=1,...,P} \{C_{u_p}\}$) capacity value of an *UE* in one of the $P$ configurations, respectively. The *PDF* and the *CDF* for the baseline and the optimal $P$-tiled layout are compared in Fig. 12(a) and Fig. 12(c), respectively. More specifically, the *PDF* distribution of the single *UE*s capacity [Fig. 12(a)] shows how the irregular *P*-shaped arrangement reduces the probability to have *UE*s served with a low throughput (e.g., $C \leq 5$ [bps/Hz]) as also confirmed by the *CDF* curves in Fig. 12(c). As a matter of fact, the CDF of $\widetilde{s}$ lies in the "green" area (i.e., the region where there is an improvement of the performance of the baseline) mainly in the bottom left part of the graph. For instance, the percentage of *UE*s served with a capacity smaller than $C \leq 5$ [bps/Hz] has been reduced from $CDF_{bsl} = 38\%$ down to $\widetilde{CDF} = 26\%$ with an improvement of $\Delta CDF = 12\%$ [Fig. 12(*c*)]. It is worthwhile pointing out that such a result (i.e. the improvement of the throughput for the poorly served *UE*s) is highly desired and it is a key-target when designing innovative *PA*s for *BS*s to enhance the *QoS* and the user-experience.

The same experiment carried out for the *2D-UMa-O* scenario has been also performed in another propagation environment where all the *UE*s were indoor within buildings (i.e., *2D-UMa-I* scenario), while keeping the same setup of the previous test case for all other parameters. Since the *PA* aperture $\mathcal{A}$ and the tile shapes [Fig. 9(*a*)] are the same, it has not been necessary to generate the tiling configurations, $\{\mathbf{s}_t;\ t = 1, ..., T\}$, but only to evaluate the corresponding capacity (12) and coverage (13) metrics. This did not imply a significant reduction of the computational cost because of the efficiency of the Algorithm-X that took 5 [sec] for generating the $T = 85926$ complete tilings, the time for the analysis of the capacity and coverage metrics being the most time-demanding step. The values of the average sum-rate capacity, $\{\widehat{C}_t;\ t = 1, ..., T\}$, are reported in Fig. 11(*b*). In this case, the number of synthesized solutions that outperform the baseline is 837 (i.e., 0.97 % of $T$) and the optimal tiling [Fig. 10(*c*)] affording the maximum capacity performance, $\widetilde{C}$, improves of $\Delta \widehat{C} = 5.44$ % (Tab. III) the baseline value ($\widetilde{C} = 140.56$ [bps/Hz] vs. $\widehat{C}_{bsl} = 133.31$ [bps/Hz] - Tab. II). Although the minimum power of the desired signal on receive, $\eta_{des}^{min}$, is lower than that of the previous scenario ($\eta_{des}^{min}\rfloor_{2D-UMa-I} = -92.92$ [dBm] vs. $\eta_{des}^{min}\rfloor_{2D-UMa-O} = -75.93$ [dBm] - Tab. II), the optimized layout still fulfils the coverage condition as indicated by the plot in Fig. 11(*d*), while the average sum-rate capacity is higher ($\widetilde{C}\big\rfloor_{2D-UMa-I} = 140.56$ [bps/Hz] vs. $\widetilde{C}\big\rfloor_{2D-UMa-O} = 130.38$ [bps/Hz] - Tab. II). This is evident from Figs. 13(*a*)-13(*b*) that plot the normalized values of the average (over the $P$ *UE*s configurations) power received by each $\chi$-polarized ($\chi \in \{V, H\}$) *RX* port of the *UE*s terminals for a given $\psi$-polarized ($\psi \in \{V, H\}$) *TX* port of the optimized irregular tiling [Fig. 13(*a*)] and of the baseline [Fig. 13(*b*)]. Since the diagonal values are related to the desired *RX* power, $\Pi_{des}$ [$\Pi_{des} \triangleq \frac{1}{P} \times \sum_{p=1}^{P} \Pi_{des}^{(\chi)}(\mathbf{r}_{u_p})$], while the others refers to the multi-user interference, $\Pi_{mui}$ [$\Pi_{mui} \triangleq \frac{1}{P} \times \sum_{p=1}^{P} \Pi_{mui}^{(\chi)}(\mathbf{r}_{u_p})$], the plots assess the effectiveness of the *ZF* method in reducing the interfering signals, while maximizing the strength of the desired one.

For comparison purposes, the difference map between the baseline values [Fig. 13(*b*)] and the *P*-shaped tiling ones [Fig. 13(*a*)] is given in Fig. 13(*c*) and, with a binary (black and white) notation, in Fig. 13(*d*). More specifically, the black and white pixels in Fig. 13(*d*) denote the links in which the irregularly-tiled *PA* provides a signal strength lower (desirable for the interfering power, $\Pi_{mui}^{(\chi)}$) or higher (desirable for the desired signal power, $\Pi_{des}^{(\chi)}$) than the baseline, respectively. One can observe that the majority of the pixels are white along the diagonal (i.e., $78.13$ %) and black outside the diagonal (i.e., $64.72$ %), thus indicating a better performance of the optimized solution. For the sake of completeness, the distributions of the *PDF* and of the *CDF* are reported in Fig. 12(*b*) and Fig. 12(*d*), respectively.

In the third example, two new sets of simulations have been carried out by considering both *2D-UMa* scenarios (i.e., outdoor and indoor), but now taking into account different tile shapes, namely the *L*-shaped hexomino tiles [Fig. 9(*b*) - $F = 1$] and jointly the *P*-shaped and the *L*-shaped hexomino tiles [Fig. 9(*c*) - $F = 2$]. The tiled layouts synthesized for the *2D-UMa-O* and the *2D-UMa-I* scenarios are given in Figs. 14(*a*)-14(*b*) and Figs. 14(*c*)-14(*d*) for the *L*-shaped and the *P*- plus *L*-shaped tiles, respectively, the total number of complete configurations being equal to $T\rfloor_L = 3656$ and $T\rfloor_{P+L} = 81986$, respectively. By observing the aperture tessellations, it can be noticed that in all cases the optimal configurations present vertically-oriented tiles, namely *P*- or *L*-shaped hexominoes with the longer segment placed along the *z*/vertical direction. Indeed, since the *PA* of the *BS* is located $25$ [m] above the *UE*s, which are on the ground, there is no need to perform a wide vertical scanning, but mainly a horizontal one, and vertically-oriented clusters are more suitable to control the shape of the beams along the azimuth plane. As for the average sum-rate capacity, the optimal irregular *L*-shaped tiles arrangement yields a performance improvement, with respect to the baseline, of $\Delta \widehat{C} = 9.83$ % for the *2D-UMa-O* scenario and of $\Delta \widehat{C} = 3.93$ % for the *2D-UMa-I* one (Tab. III). When jointly using *P*-shaped and *L*-shaped hexomino tiles, such an improvement grows up to $\Delta \widehat{C} = 14.14$ % and $\Delta \widehat{C} = 6.29$ % for the outdoor and indoor cases, respectively (Tab. III), also outperforming the *P*-shaped tiling solution in Fig. 10, but only for $2.15$ % (*2D-UMa-O*) and $0.85$ % (*2D-UMa-I*), despite the use of more types of tiles (i.e., $F = 2$). Moreover, the *CDF* in Figs. 14(*e*)-14(*f*) indicates that the irregular clustering with the three different tile sets in Fig. 9 performs similarly since the corresponding curves almost overlap. Concerning the coverage, the link-connectivity condition (13) holds true for all the optimal solutions in Fig. 14 as quantitatively confirmed by the values in Tab. III.

The next two examples deal with a different distribution of the *UE*s within the cell, which now are randomly-placed on different levels, instead of being all placed at the ground level, to simulate their presence inside the floors of the buildings. According to the *3GPP* guidelines [39], the height of the $u$-th ($u = 1, ..., U$) *UE* in the $p$-th ($p = 1, ..., P$) configuration has been determined according to the following rule

$$h_{UE, u_p} = 3 \times (n_{floor} - 1) + 1.5, \quad (20)$$

where $1 \leq n_{floor} \leq \Omega$ and $4 \leq \Omega \leq 8$ are uniformly-distributed random integer numbers. The irregular *PA* has been then designed by considering the *3D-UMa* scenario shown in Fig. 15(*a*) and the indoor propagation model (*3D-UMa-I*), while the *P*-shaped tiles of Fig. 9(*a*) have been used. The optimal tessellation synthesized with the proposed *CDTA* method [Fig. 15(*c*)] turns out to be different from that of the *2D-UMa-I* scenario [Fig. 10(*c*)], even though, it is still composed by only vertically-oriented tiles. As for the comparison with the baseline, the optimized layout enhances the capacity performance of $\Delta \widehat{C} = 7.59$ % (Tab. III) since $\widetilde{C} = 130.76$ [bps/Hz], while $\widehat{C}_{bsl} = 121.54$ [bps/Hz] (Tab. IV). Such an improvement, while fulfilling the coverage condition (Tab. IV), is confirmed by the *CDF* curves in Fig. 15(*e*).

The last example investigates a communication scenario (*3D-UMi-I*) characterized by a smaller *Urban-Micro* (*UMi*) cell (i.e., $ISD = 200$ [m]) with a *BS* closer to the ground floor (i.e., $h_{BS} = 10$ [m] [39]) [Fig. 15(*b*)]. The hexagonal cell being smaller, the number of *UE*s configurations has been reduced ($P = 150$), while all the remaining descriptive parameters have been kept equal to the *3D-UMa-I* case. The best optimized arrangement of the tiles over the aperture and the corresponding *CDF* curve are shown in Fig. 15(*d*) and Fig. 15(*f*), respectively. Unlike the previous array layouts, horizontally-oriented *P*-shaped tiles are now present [Fig. 15(*d*)] since here there are *UE*s located up to 12.5 [m] above the level of the *BS* and a vertical beam scanning wider than that for the *3D-UMa-I* one is now needed. For completeness, the capacity and the coverage indexes of the optimally-synthesized *PA* are reported in Tab. IV. As it can be inferred, there is a $\Delta \widehat{C} = 9.82$ % (Tab. III) improvement of the baseline performance with a full coverage of the *UMi* cell (i.e., $\eta_{des}^{min} > \Pi_{th}$ being $\eta_{des}^{min} = -95.67 \, [dBm]$ - Tab. IV).

Finally, the average performance of each synthesized irregular *PA* solution over different scenarios have been evaluated to take into account that, generally, a *BS* model is used in different scenarios of the same class of environments (e.g., rural, metropolitan, etc ...). Towards this end, the optimal layouts derived for each specific working condition [i.e., *2D-UMa-O* - Fig. 10(*b*), Fig. 14(*a*), and Fig. 14(*c*); *2D-UMa-I* - Fig. 10(*c*), Fig. 14(*b*), and Fig. 14(*d*); *3D-UMa-I* - Fig. 15(*c*); *3D-UMi-I* - Fig. 15(*d*)] have been tested in the other/different scenarios taken into account in this work. The results in terms of variations of the average sum-rate capacity $\Delta \widehat{C}$ with respect to the baseline are reported in Tab. III. In two cases, namely the *PA*s of the *2D-UMa-I* scenario when using *L*-shaped and jointly *P*- and *L*-shaped tiles, there is always an improvement with respect to the baseline. In the overall, 38 out of 48 checks (80 %) [Tab. III] are positive for the irregularly-clustered *PA*s with capacity enhancements up to $+36.29$ % (Tab. III), while only in few cases (i.e., 10 out of $48 \to 20$ %) the baseline works better and there is a degradation of at most $-4.94$ % whether installing an irregular clustered *PA*. The most critical condition turns out to be the case of the *PA* optimized for the *UMi* cell when used in the *UMa* scenarios. This result suggests the need of a customized solution for the micro cells, while generally the irregular clustering turns out to be robust and effective.

# 5 Conclusions

A novel capacity-driven method for the synthesis of aperiodically-tiled *PA*s in *MU-MIMO* communications has been presented. In order to maximize the average sum-rate capacity of the *PA* in downlink, while assuring the coverage of the *BS* cell, the proposed approach has been developed by integrating a customized version of the Algorithm-X, to generate the whole set of complete tessellations of the *PA* aperture, and the Zero-Forcing technique, to compute the sub-array excitations.

From a methodological viewpoint and to the best of the authors' knowledge, the main novelties of this research work can be summarized as follows:

- the robust synthesis of irregular tiled arrays for wireless communications able to maximize the channel capacity, while guaranteeing the coverage within the whole communication cell, thus the link connections between the *BS* and *UE*s;

- the definition of a nested approach for the capacity-based synthesis of unconventional *PA*s based on a tiling algorithm able, for any tiles alphabet, to generate the whole set of complete tessellations of the *PA* aperture.

From the numerical assessment, which has been carried out by considering real radiating elements and *3GPP*-recommended propagation scenarios, the following outcomes can be drawn:

- there is a huge number of irregularly-tiled layouts that outperform the baseline so that the antenna engineers have at disposal a wide set of options/choices for taking into account other constraints and requirements including costs and manufacturing complexity;

- although the optimal tiling is scenario-dependent, the irregular layouts synthesized for a specific scenario turns out to be generally advantageous with respect to the baseline in the majority of the other propagation environments;

- the relative positions between the *BS* and the *UE*s have an impact on the orientation of the tiles in the optimal clustering of the aperture. Dealing with small cells (i.e., *UMi*) with the *BS* placed also below the *UE*s level, the optimal tessellations count horizontally-placed tiles, while only vertically-oriented clusters are present in larger cells (i.e., *UMa*) when the *BS* is above the *UE*s;

- all the optimized aperiodic tiled arrays fulfil the coverage condition, thus they can be considered as reliable solutions for real communication systems.

Future research activities, beyond the scope of this paper, will consider the design of *PA*s having different aperture shapes (e.g., circular and hexagonal) or conformal layouts (e.g., spherical, cylindrical) and the use of other tiles as well as the extension to other frequency bands of interest for future-generation communications.

# Acknowledgements

This work benefited from the networking activities carried out within the Project Project "CYBER-PHYSICAL ELECTROMAGNETIC VISION: Context-Aware Electromagnetic Sensing and Smart Reaction (EMvisioning)" (Grant no. 2017HZJXSZ)" funded by the Italian Ministry of Education, University, and Research under the PRIN2017 Program (CUP: E64I19002530001). Moreover, it benefited from the networking activities carried out within the Project "SPEED" (Grant No. 61721001) funded by National Science Foundation of China under the Chang-Jiang Visiting Professorship Program, the Project 'Inversion Design Method of Structural Factors of Conformal Load-bearing Antenna Structure based on Desired EM Performance Interval' (Grant no. 2017HZJXSZ) funded by the National Natural Science Foundation of China, and the Project 'Research on Uncertainty Factors and Propagation Mechanism of Conformal Loab-bearing Antenna Structure' (Grant No. 2021JZD-003) funded by the Department of Science and Technology of Shaanxi Province within the Program Natural Science Basic Research Plan in Shaanxi Province. A. Massa wishes to thank E. Vico for her never-ending inspiration, support, guidance, and help.

# FIGURE CAPTIONS

- **Figure 1.** Sketch of a mobile communication scenario with hexagonal cells.

- **Figure 2.** Pictorial representation of the *PA* geometry.

- **Figure 3.** Clustered configurations and entries of the corresponding sub-array aggregation vector **s** of (*a*) a baseline/regularly-tiled array composed by vertical sub-arrays and of (*b*) an irregularly-tiled array with *P*-shaped tiles.

- **Figure 4.** *Tiling Problem* ($M = 3$, $N = 2$, $K = 7$) - Set of admissible positions and orientations of a domino tile within the rectangular aperture $\mathcal{A}$ of $M \times N$ elements, $\mathcal{T} = \{\tau_1, \tau_2, \tau_3, \tau_4, \tau_5, \tau_6, \tau_7\}$.

- **Figure 5.** *Tiling Problem* ($M = 3$, $N = 2$, $K = 7$) - Examples of (*a*)(*c*) the layouts and of (*b*)(*d*) the corresponding incident matrix **L** for (*a*) an admissible/non-overlapping tiling (i.e., $\mathbf{s} = \{1; 1; 2; 3; 3; 2\}$, $\widehat{\mathcal{T}} = \{\tau_{\hat{1}} = \tau_1, \tau_{\hat{2}} = \tau_3, \tau_{\hat{3}} = \tau_7\}$) and for (*c*) a forbidden/overlapping tiling.

- **Figure 6.** *Tiling Problem* ($M = 3$, $N = 2$, $K = 7$) - Sketch of (*a*)(*c*)(*e*) the operations performed on the incident matrix **L** and of (*b*)(*d*)(*f*) the corresponding tiles placement according to the Algorithm-X.

- **Figure 7.** *Tiling Problem* ($M = 3$, $N = 2$, $K = 7$) - Sketch of (*a*)(*b*)(*c*) the operations on the incident matrix **L** towards the generation of the successive tiles arrangement and (*d*) placement of the first tile for the new $(t + 1)$ tiled solution.

- **Figure 8.** *Numerical Assessment* - Sketch of (*a*) the array element and of (*b*) the baseline regularly-clustered array configuration.

- **Figure 9.** *Numerical Assessment* - Set of (*a*) *P*-shaped, (*b*) *L*-shaped and (*c*) (*P+L*)-shaped tiles.

- **Figure 10.** *Numerical Assessment* (*P-shaped Tiles*, *2D-UMa*, $M = 8$, $N = 12$, $d_y = 0.5\lambda$, $d_z = 0.7\lambda$, $Q = U = 16$, $P = 200$) - Picture of (*a*) the complete set of $P \times U$

*UE* positions within the *3GPP 2D-UMa* hexagonal cell scenario and of (*b*)(*c*) the optimal irregular tiling for (*b*) the outdoor (*2D-UMa-O*) and (*c*) the outdoor-to-indoor (*2D-UMa-I*) scenarios.

- **Figure 11.** *Numerical Assessment (P-shaped Tiles, 2D-UMa, $M = 8$, $N = 12$, $d_y = 0.5\lambda$, $d_z = 0.7\lambda$, $Q = U = 16$, $P = 200$)* - Behavior of (*a*)(*b*) the average sum-rate capacity $\widehat{C}$ of the $T$ admissible tilings and of the baseline along with (*c*)(*d*) the level of the *RX* power of the desired signal received at each *UE* port when adopting the optimal irregular tiling in (*a*)(*c*) the outdoor (*2D-UMa-O*) and (*b*)(*d*) the outdoor-to-indoor (*2D-UMa-I*) propagation scenario.

- **Figure 12.** *Numerical Assessment (P-shaped Tiles, 2D-UMa, $M = 8$, $N = 12$, $d_y = 0.5\lambda$, $d_z = 0.7\lambda$, $Q = U = 16$, $P = 200$)* - Behavior of (*a*)(*b*) the *PDF* and of (*c*)(*d*) the *CDF* when adopting the optimal irregular tiling or the baseline configuration in (*a*)(*c*) the outdoor (*2D-UMa-O*) and (*b*)(*d*) the outdoor-to-indoor (*2D-UMa-I*) propagation scenarios.

- **Figure 13.** *Numerical Assessment (P-shaped Tiles, 2D-UMa, $M = 8$, $N = 12$, $d_y = 0.5\lambda$, $d_z = 0.7\lambda$, $Q = U = 16$, $P = 200$)* - Plot of (*a*)(*b*) the entries of the *RX* power matrix when considering $U = 16$ users and a *PA* with $Q = 16$ sub-arrays arranged according to (*a*) the optimal irregular tiling or (*b*) as in the baseline configuration. Plot of the difference map between the entries of the *RX* power matrix of the baseline and of the *P*-shaped tiling in (*c*) color and (*d*) thresholded (black/white).

- **Figure 14.** *Numerical Assessment (2D-UMa, $M = 8$, $N = 12$, $d_y = 0.5\lambda$, $d_z = 0.7\lambda$, $Q = U = 16$, $P = 200$)* - Picture of (*a*)(*b*)(*c*)(*d*) the optimal irregular tiling and (*e*)(*f*) *CDF* of both the optimal irregular tiling and the baseline configuration in (*a*)(*c*)(*e*) the outdoor (*2D-UMa-O*) and (*b*)(*d*)(*f*) the outdoor-to-indoor (*2D-UMa-I*) propagation scenarios.

- **Figure 15.** *Numerical Assessment (P-shaped Tiles, $M = 8$, $N = 12$, $d_y = 0.5\lambda$, $d_z = 0.7\lambda$, $Q = U = 16$)* - Picture of (*a*)(*b*) the $P \times U$ *UE* positions of the *3GPP* (*a*) *3D-UMa* ($P = 200$) and (*b*) *3D-UMi* ($P = 150$) hexagonal cell scenarios. Sketch of (*c*)(*d*) the optimal irregular tiling and plot of (*e*)(*f*) the *CDF* of both the optimal irregular tiling and the baseline configuration for the outdoor-to-indoor propagation scenario and (*a*)(*c*)(*e*) the *UMa* (*3D-UMa-I*) and (*b*)(*d*)(*f*) the *UMi* (*3D-UMi-I*) hexagonal cells.

## TABLE CAPTIONS

- **Table I.** *3GPP* scenario parameters setup.

- **Table II.** *Numerical Assessment* (*2D-UMa*, $M = 8$, $N = 12$, $d_y = 0.5\lambda$, $d_z = 0.7\lambda$, $Q = U = 16$, $P = 200$) - Statistics of the sum-rate capacity, $C$, and of the minimum power of the desired signal $\eta_{des}^{(a)}$ ($a = 1, ..., A$; $A \triangleq 2 \times U$).

- **Table III.** *Numerical Assessment* ($M = 8$, $N = 12$, $d_y = 0.5\lambda$, $d_z = 0.7\lambda$, $Q = U = 16$) - Average sum-rate capacity variations, $\Delta\widehat{C}$.

- **Table IV.** *Numerical Assessment* ($M = 8$, $N = 12$, $d_y = 0.5\lambda$, $d_z = 0.7\lambda$, $Q = U = 16$) - Statistics of the sum-rate capacity, $C$, and of the minimum power of the desired signal $\eta_{des}^{(a)}$ ($a = 1, ..., A$; $A \triangleq 2 \times U$).

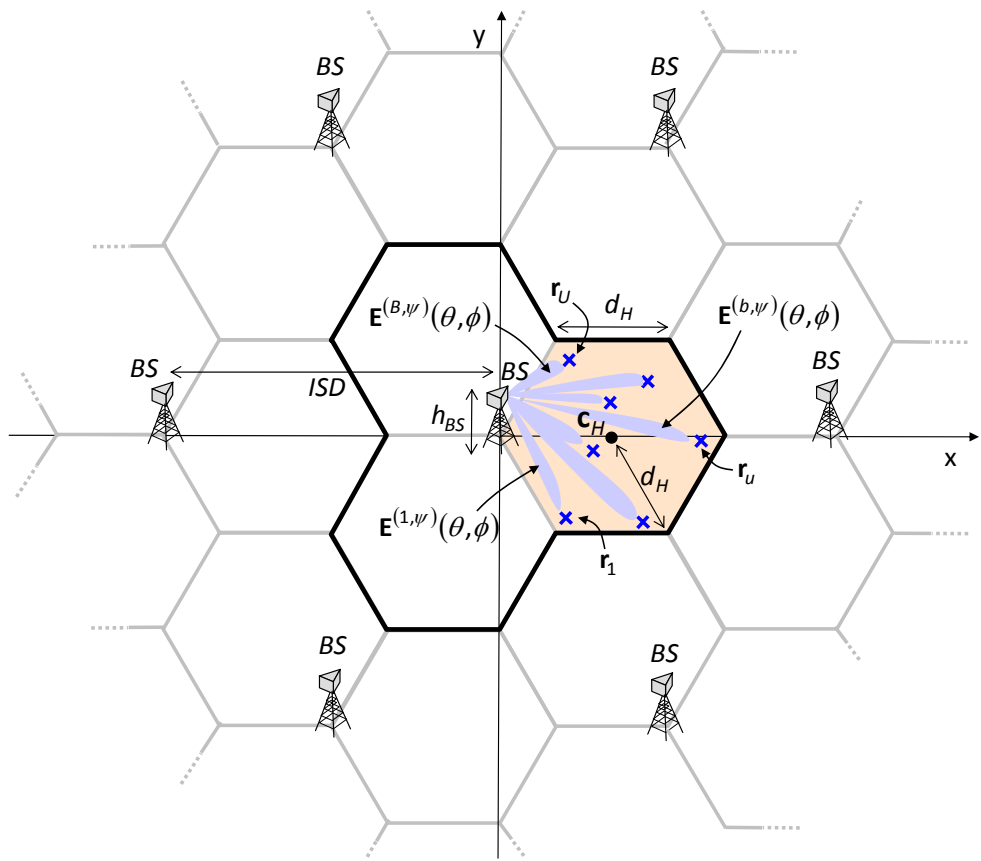

**Fig. 1** - N. Anselmi *et al.*, "Optimal Capacity-Driven Design ..."

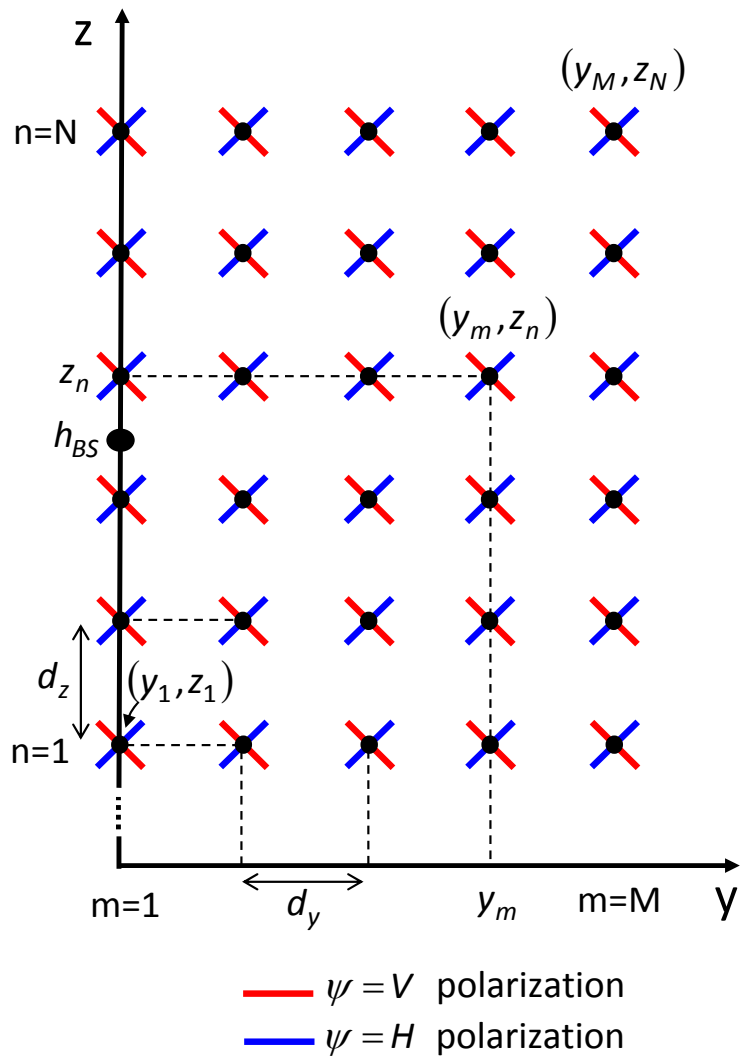

**Fig. 2** - N. Anselmi *et al.*, "Optimal Capacity-Driven Design ..."

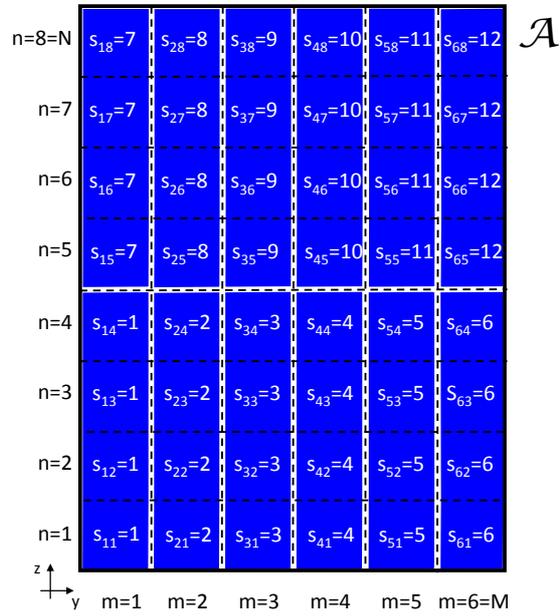

(*a*)

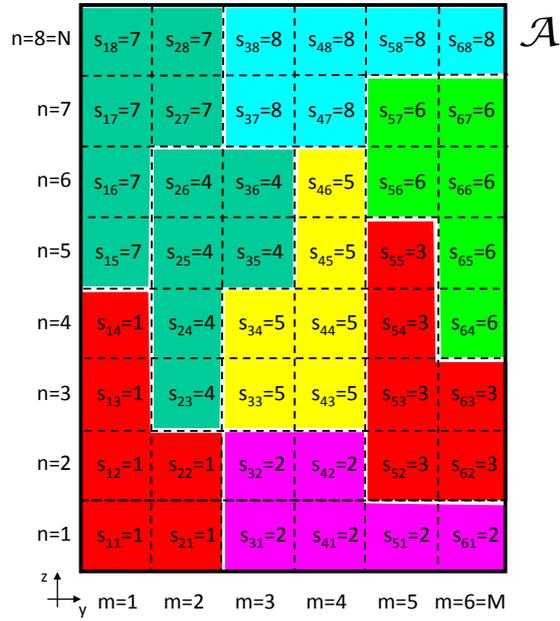

(*b*)

**Fig. 3 -  N. Anselmi *et al.*, "**Optimal Capacity-Driven Design ..."

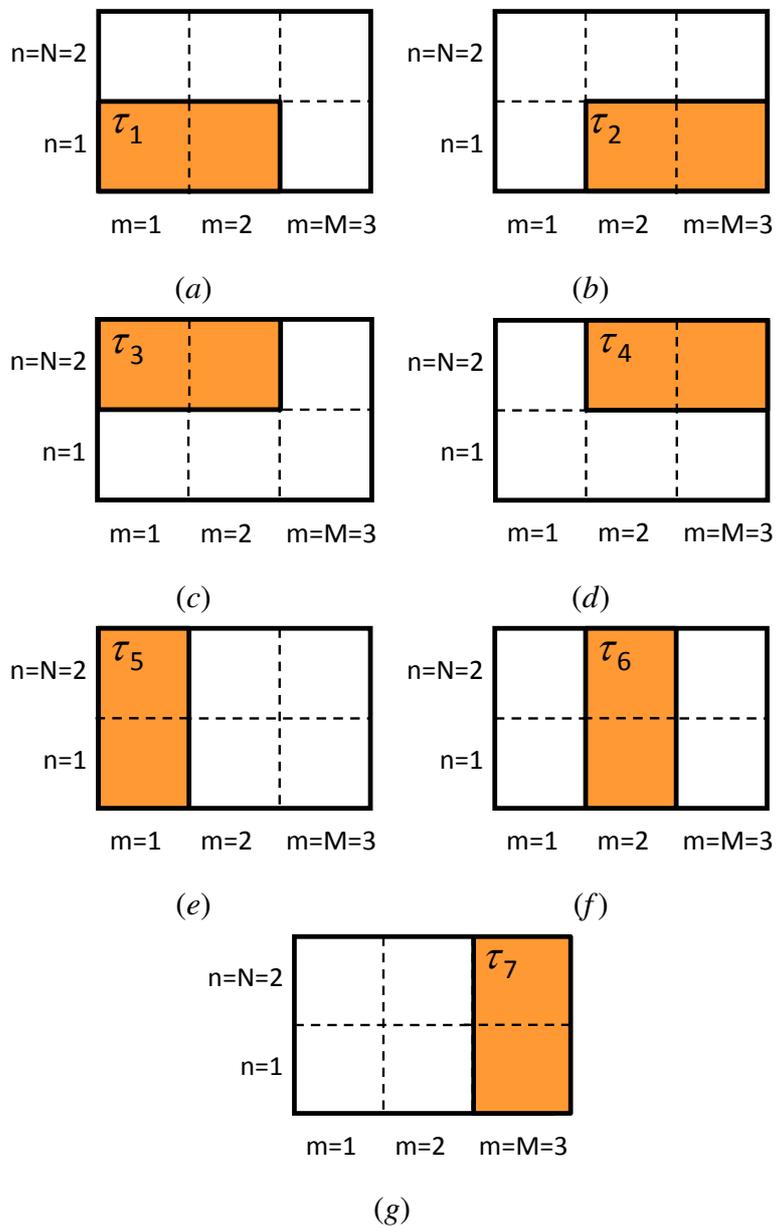

**Fig. 4** - N. Anselmi *et al.*, "Optimal Capacity-Driven Design ..."

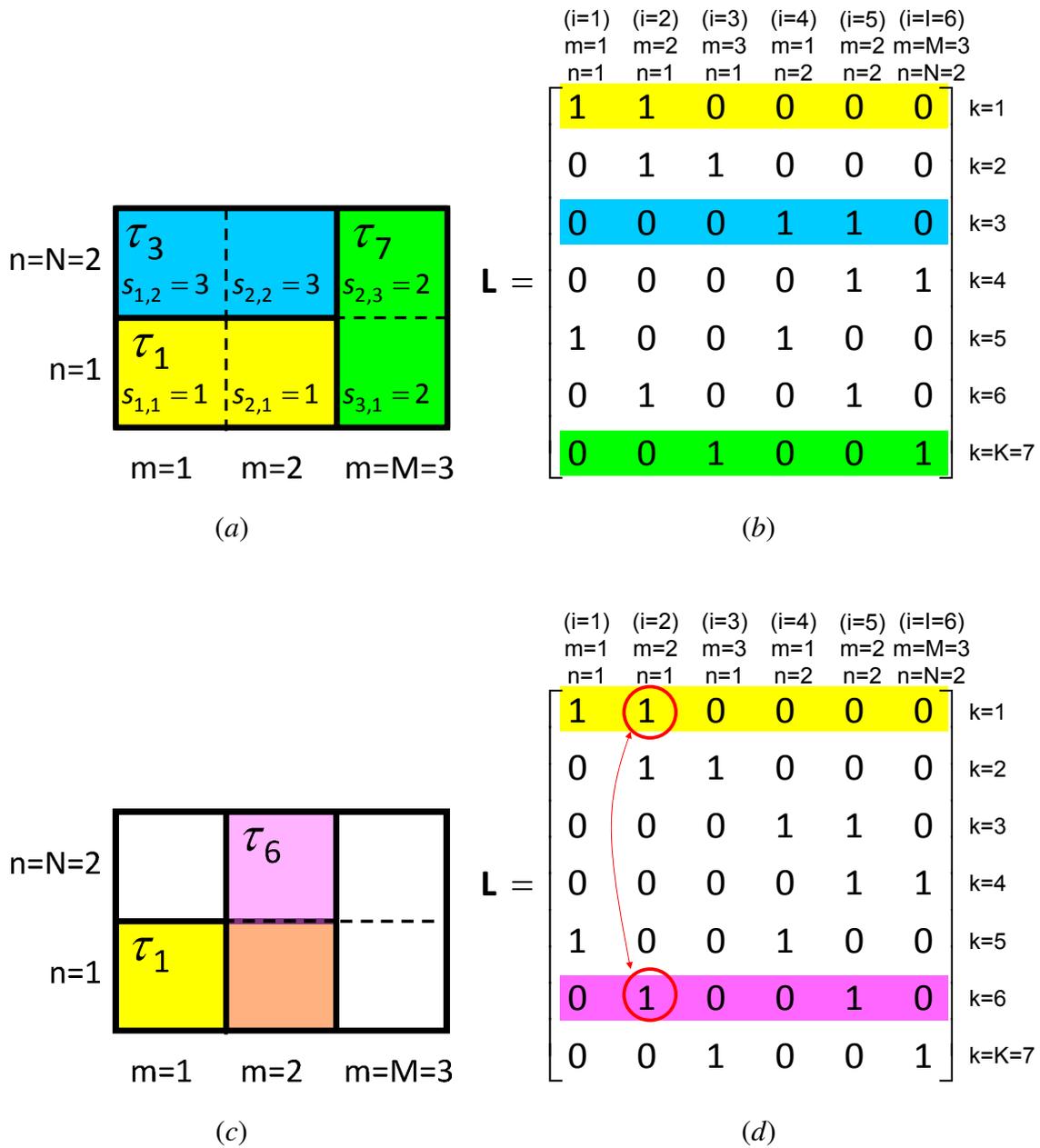

**Fig. 5 -** N. Anselmi *et al.*, "Optimal Capacity-Driven Design ..."

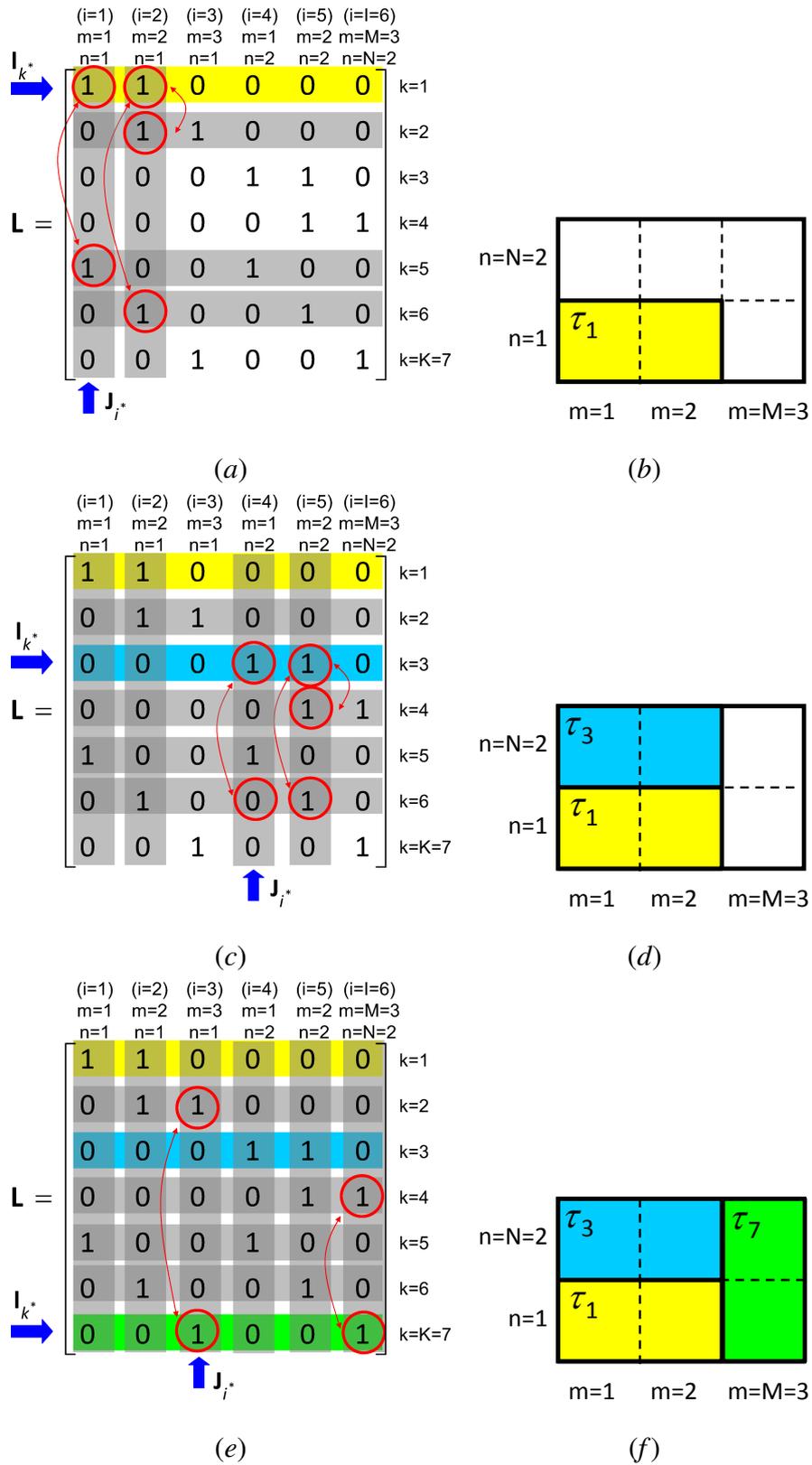

**Fig. 6 -** N. Anselmi *et al.*, "Optimal Capacity-Driven Design ..."

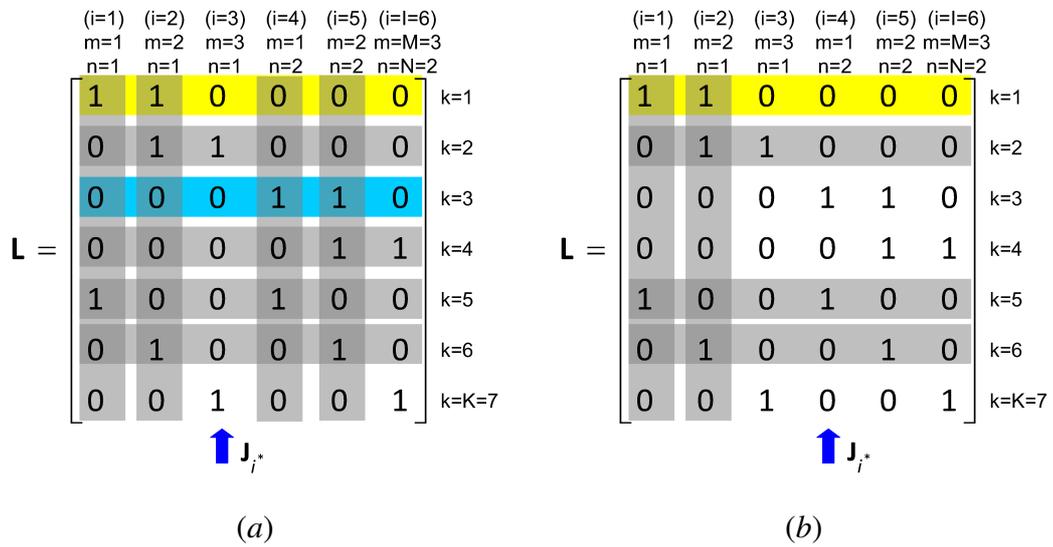

(a)   (b)

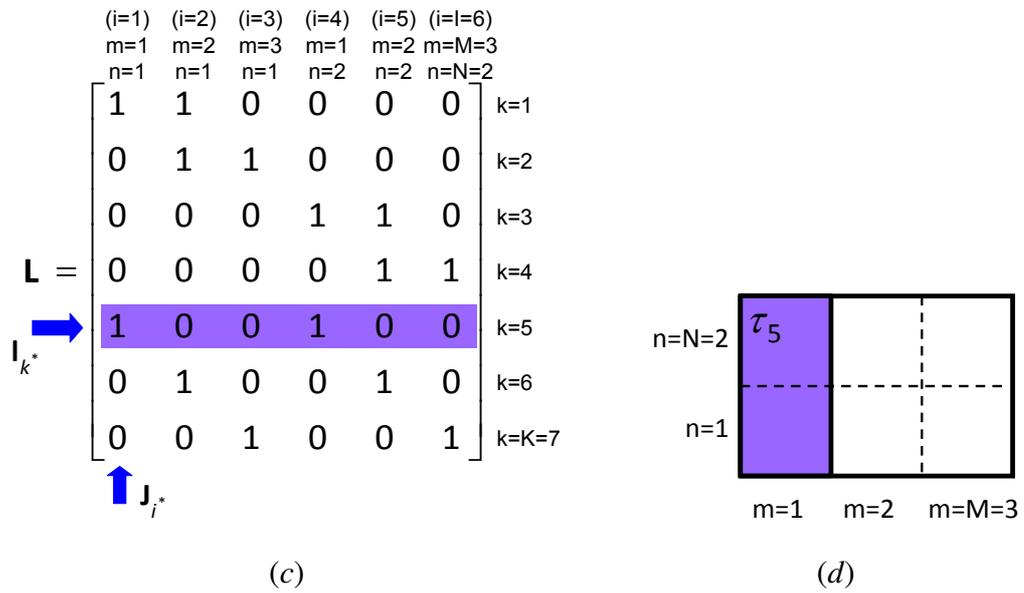

(c)   (d)

**Fig. 7 -** N. Anselmi *et al.*, "Optimal Capacity-Driven Design ..."

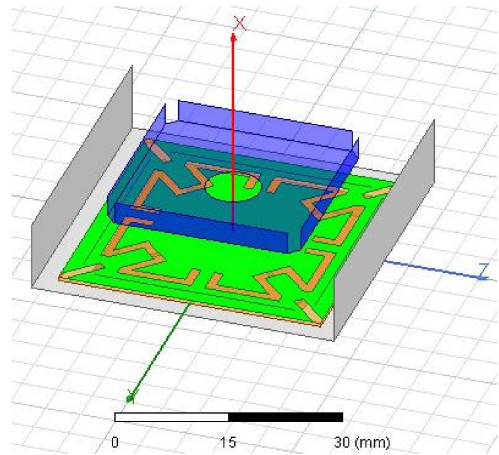

(*a*)

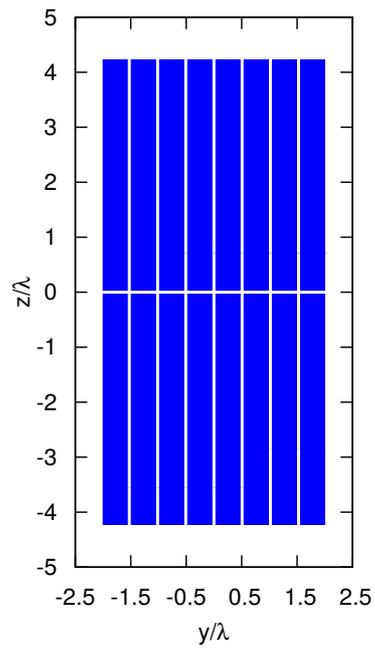

(*b*)

**Fig. 8 - N. Anselmi *et al.*,** "Optimal Capacity-Driven Design ..."

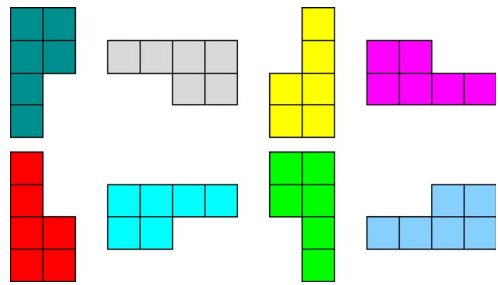

(a)

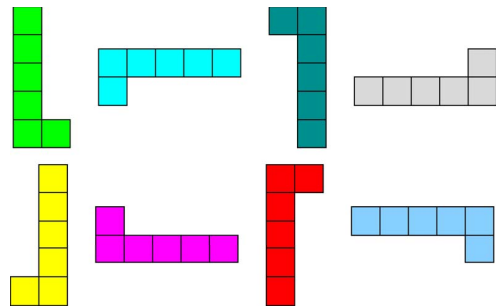

(b)

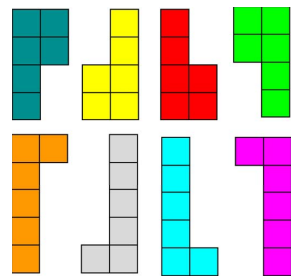

(c)

**Fig. 9 -** N. Anselmi *et al.*, "Optimal Capacity-Driven Design ..."

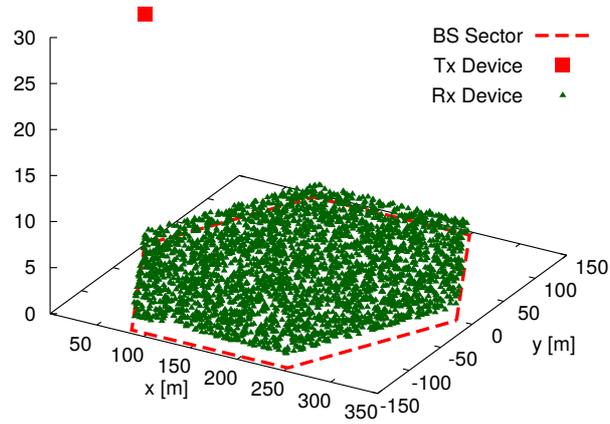

(*a*)

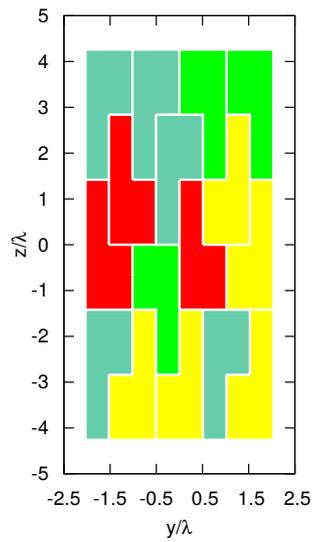

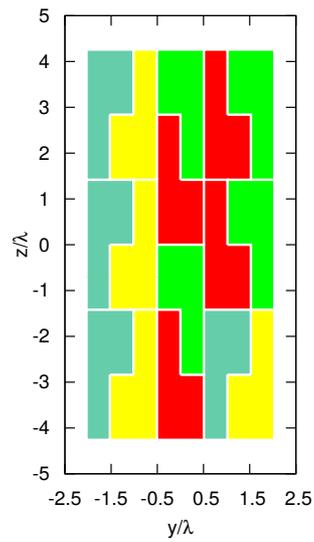

(*b*)                          (*c*)

**Fig. 10 - N. Anselmi *et al.*, "Optimal Capacity-Driven Design ..."**

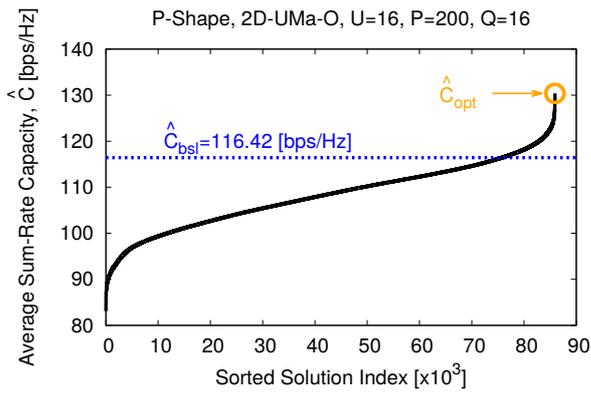
(a)

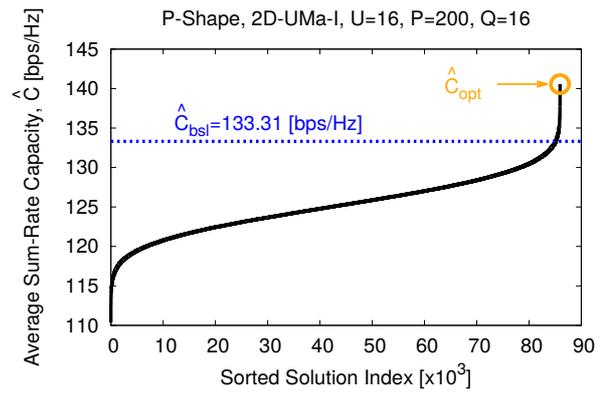
(b)

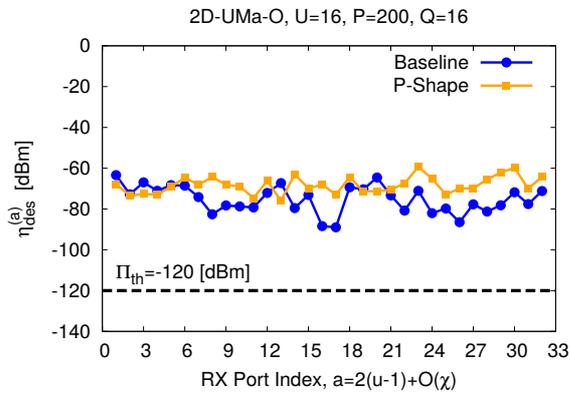
(c)

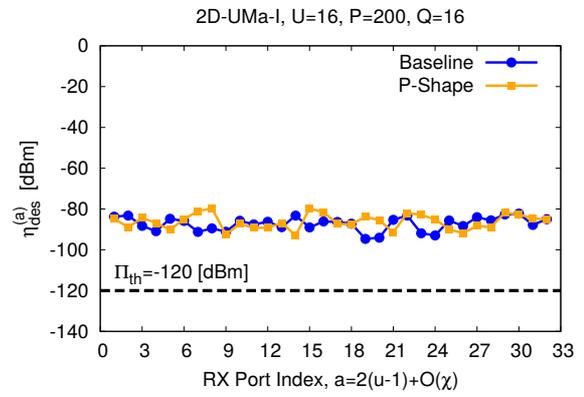
(d)

**Fig. 11 -** N. Anselmi *et al.*, "Optimal Capacity-Driven Design ..."

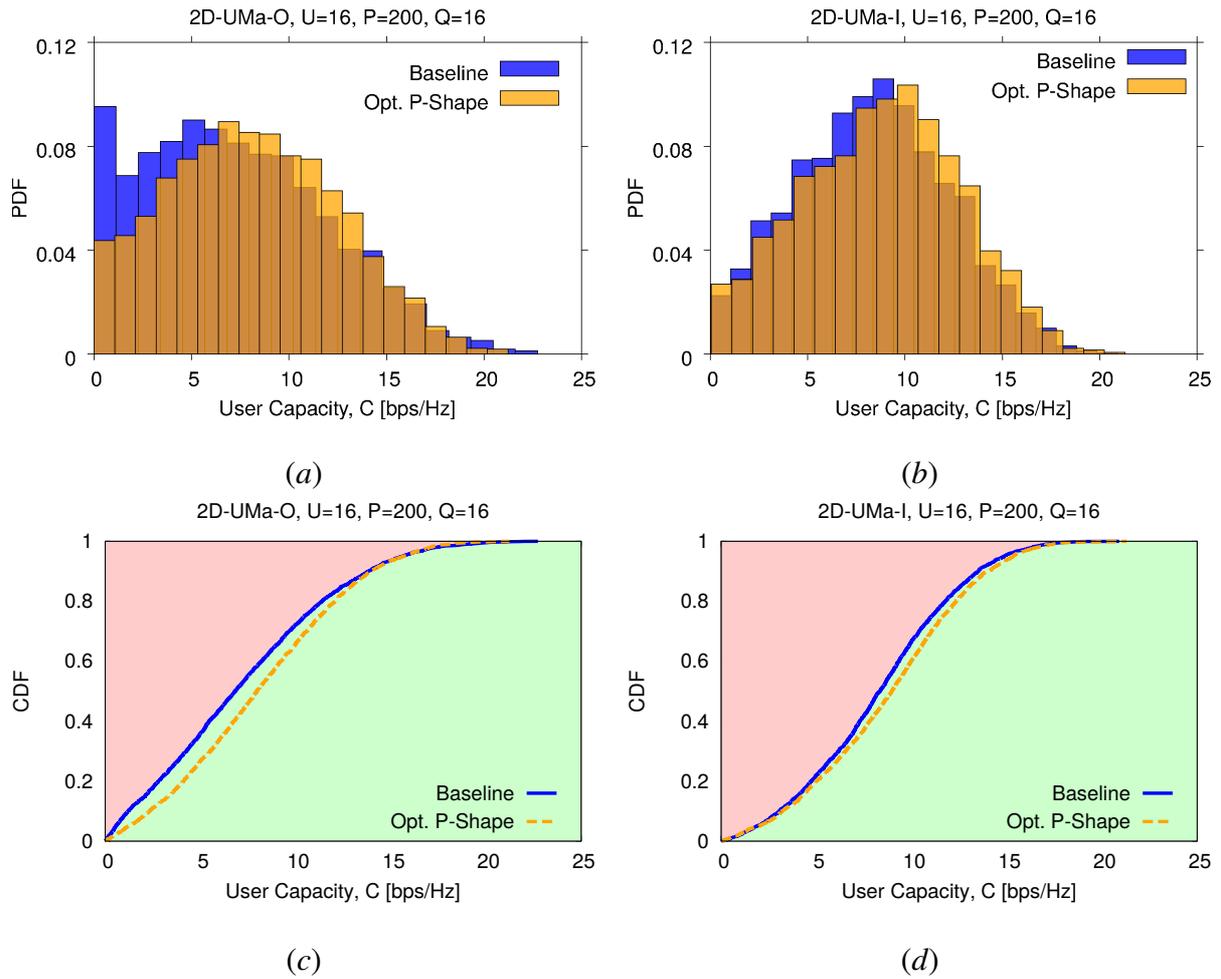

**Fig. 12 -** N. Anselmi *et al.*, "Optimal Capacity-Driven Design ..."

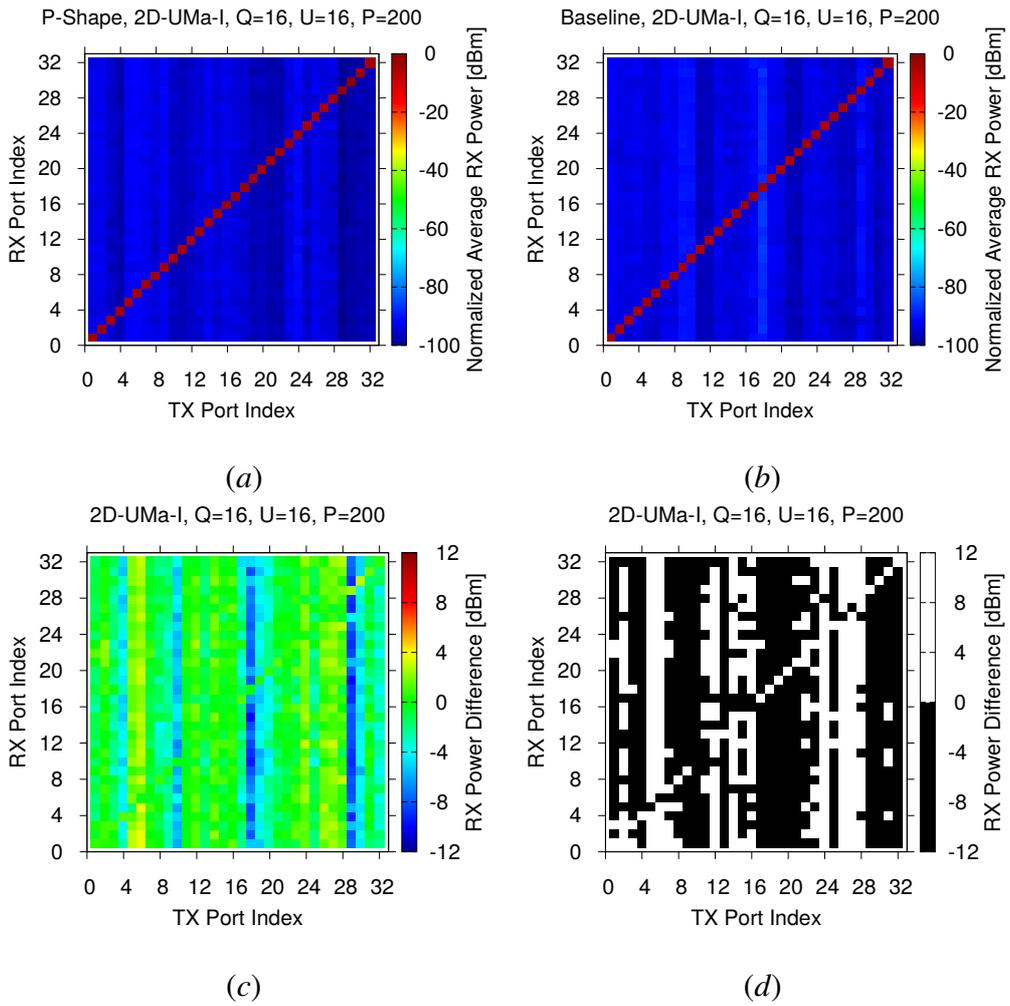

**Fig. 13 - N. Anselmi *et al.*, "**Optimal Capacity-Driven Design ..."

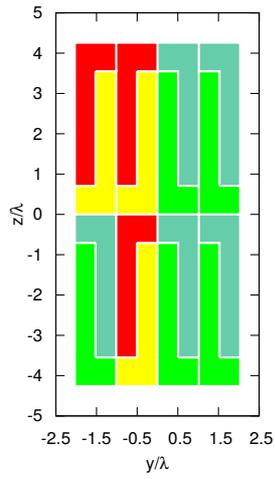

(a)

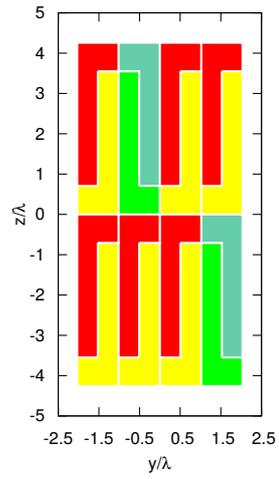

(b)

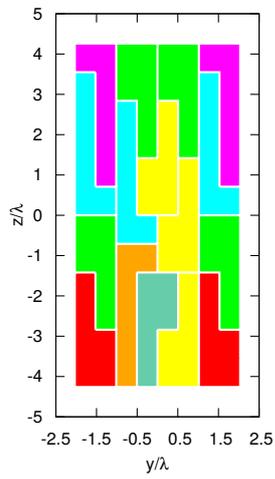

(c)

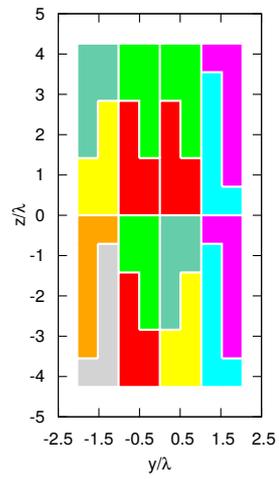

(d)

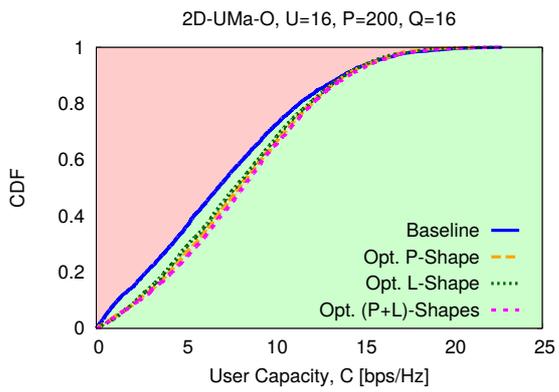

(e)

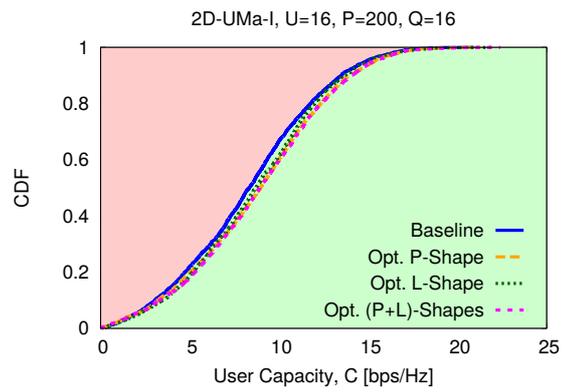

(f)

**Fig. 14 -** **N. Anselmi** *et al.***,** "Optimal Capacity-Driven Design ..."

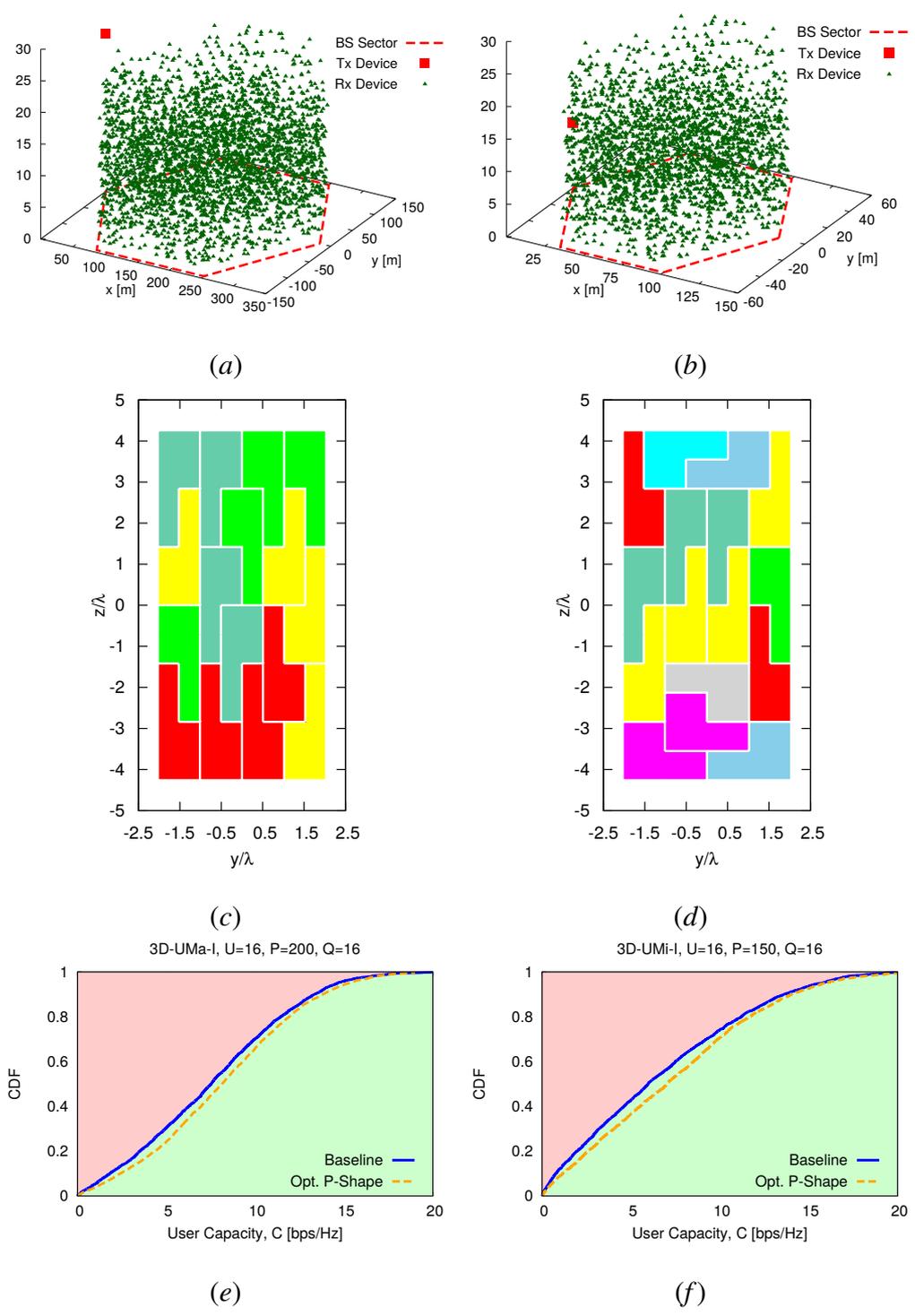

**Fig. 15** - N. Anselmi *et al.*, "Optimal Capacity-Driven Design ..."

|  | *UMa* | *UMi* |
|---|---|---|
| $h_{BS}$ [m] | 25.0 | 10.0 |
| $ISD$ [m] | 500 | 200 |
| $h_{UE}$ (2D) [m] | 1.5 | |
| $h_{UE}$ (3D) [m] | $3 \times (n_{floor} - 1) + 1.5$ | |
| $\Psi$ [dBm] | 43 | |

**Tab. I - N. Anselmi** *et al.***,** "Optimal Capacity-Driven Design ..."

|  | Sum-Rate Capacity, $C$ [bps/Hz] | | | | Min. RX Power, $\eta_{des}^{(a)}$ [dBm] | | | |
|---|---|---|---|---|---|---|---|---|
|  | Min. | Max. | Avg. | Var. | Min. | Max. | Avg. | Var. |
|  | 2D-UMa, Outdoor | | | | | | | |
| *Baseline* | 25.66 | 212.05 | 116.42 | $1.52 \times 10^3$ | $-88.98$ | $-63.38$ | $-75.28$ | 43.05 |
| *P-Shape* | 6.10 | 211.53 | 130.38 | $1.14 \times 10^3$ | $-75.93$ | $-59.38$ | $-68.27$ | 17.05 |
| *L-Shape* | 24.91 | 215.73 | 127.86 | $1.24 \times 10^3$ | $-90.86$ | $-66.92$ | $-78.79$ | 43.97 |
| *(P+L)-Shapes* | 7.17 | 199.14 | 132.39 | $1.23 \times 10^3$ | $-77.81$ | $-56.43$ | $-65.95$ | 28.02 |
|  | 2D-UMa, Indoor | | | | | | | |
| *Baseline* | 39.38 | 203.67 | 133.31 | $1.27 \times 10^3$ | $-94.68$ | $-82.23$ | $-87.26$ | 11.22 |
| *P-Shape* | 14.99 | 204.68 | 140.56 | $1.35 \times 10^3$ | $-92.92$ | $-79.60$ | $-86.21$ | 13.40 |
| *L-Shape* | 42.85 | 213.98 | 138.55 | $1.17 \times 10^3$ | $-97.73$ | $-81.34$ | $-86.87$ | 13.16 |
| *(P+L)-Shapes* | 3.49 | 216.08 | 140.85 | $1.03 \times 10^3$ | $-101.61$ | $-81.55$ | $-88.29$ | 22.51 |

**Tab. II - N. Anselmi *et al.*,** "Optimal Capacity-Driven Design ..."

| Average Sum-Rate Capacity Improvement, $\Delta \widehat{C}$ [%] | | | | | | |
|---|---|---|---|---|---|---|
| *Optimal Tiling* | *2D-UMa* | | *3D-UMa* | | *3D-UMi* | |
| | *Outdoor* | *Indoor* | *Outdoor* | *Indoor* | *Outdoor* | *Indoor* |
| *P-Shape - 2D-UMa-O* | $+11.99$ | $+3.62$ | $-0.29$ | $+3.97$ | $+36.29$ | $+1.30$ |
| *P-Shape - 2D-UMa-I* | $-2.72$ | $+5.44$ | $+0.04$ | $+1.07$ | $+26.16$ | $+1.44$ |
| *L-Shape - 2D-UMa-O* | $+9.83$ | $-2.47$ | $+3.89$ | $+4.53$ | $+17.06$ | $-2.74$ |
| *L-Shape - 2D-UMa-I* | $+5.64$ | $+3.93$ | $+5.06$ | $+2.31$ | $+14.03$ | $+4.44$ |
| *(P+L)-Shape - 2D-UMa-O* | $+14.14$ | $-1.28$ | $+4.01$ | $+2.07$ | $+26.13$ | $+0.62$ |
| *(P+L)-Shape - 2D-UMa-I* | $+1.11$ | $+6.29$ | $+1.13$ | $+4.02$ | $+18.14$ | $+1.24$ |
| *P-Shape - 3D-UMa-I* | $+6.91$ | $-4.94$ | $+7.16$ | $+7.59$ | $+27.92$ | $+2.38$ |
| *P-Shape - 3D-UMi-I* | $-1.67$ | $-3.73$ | $-4.36$ | $-4.01$ | $+22.87$ | $+9.82$ |

**Tab. III** - N. Anselmi *et al.*, "Optimal Capacity-Driven Design ..."

|  | Sum-Rate Capacity, $C$ [bps/Hz] | | | | Min. RX Power, $\eta_{des}^{(a)}$ [dBm] | | | |
|---|---|---|---|---|---|---|---|---|
|  | *Min.* | *Max.* | *Avg.* | *Var.* | *Min.* | *Max.* | *Avg.* | *Var.* |
|  | *3D-UMa, Indoor* | | | | | | | |
| *Baseline* | 14.58 | 199.48 | 121.54 | $1.72 \times 10^3$ | $-94.75$ | $-78.19$ | $-85.01$ | 13.76 |
| *P-Shape* | 18.61 | 199.67 | 130.76 | $1.39 \times 10^3$ | $-105.62$ | $-79.77$ | $-88.64$ | 37.69 |
|  | *3D-UMi, Indoor* | | | | | | | |
| *Baseline* | 17.35 | 179.13 | 107.29 | $1.27 \times 10^3$ | $-94.70$ | $-72.78$ | $-81.24$ | 28.46 |
| *P-Shape* | 17.15 | 188.16 | 117.83 | $1.13 \times 10^3$ | $-95.67$ | $-80.05$ | $-85.80$ | 15.16 |

**Tab. IV - N. Anselmi *et al.*, "Optimal Capacity-Driven Design ..."**